\documentclass[12pt,draftcls,journal,letterpaper,onecolumn,twoside]{IEEEtran}
\usepackage{algorithm}
\usepackage{algpseudocode}
\usepackage{amsfonts}
\usepackage{bm}
\usepackage{amsmath}
\usepackage{cite}
\usepackage{float}
\usepackage{amssymb}
\usepackage{enumerate}
\usepackage{color,xcolor}
\usepackage{graphicx}
\usepackage{multirow,tabularx}
\usepackage{subfigure}
\usepackage{hyperref}

\makeatletter

\newcommand{\Rmnum}[1]{\expandafter\@slowromancap\romannumeral #1@}
\makeatother

\usepackage{bookmark}

\ifCLASSINFOpdf

\else

\fi

\begin{document}

\title{{Capacity-Achieving Iterative LMMSE Detection for MIMO-NOMA Systems}}

\author{\IEEEauthorblockN{Lei Liu, \emph{Student Member, IEEE,} Chau Yuen, \emph{Senior Member, IEEE,}\\ Yong Liang Guan, \emph{Senior Member, IEEE,} and Ying Li, \emph{Member, IEEE}}

\thanks{Lei Liu and Ying Li are with the State Key Lab of Integrated Services Networks, Xidian University, Xian, 710071, China (e-mail: lliu\_0@stu.xidian.edu.cn, yli@mail.xidian.edu.cn).}
\thanks{Chau Yuen is with the Singapore University of Technology and Design, Singapore (e-mail: yuenchau@sutd.edu.sg).}
\thanks{Yong Liang Guan is with the School of Electrical and Electronic Engineering, Nanyang Technological University, Singapore (e-mail: eylguan@ntu.edu.sg).}
\thanks{This paper was presented in part at 2016 IEEE International Conference on Communications.}
}


\maketitle

\vspace{-1.0cm}
\begin{abstract}
This paper considers a \emph{Iterative Linear Minimum Mean Square Error} (LMMSE) detection for the uplink \emph{Multiuser Multiple-Input and Multiple-Output} (MU-MIMO) systems with \emph{Non-Orthogonal Multiple Access} (NOMA), in which all the users interfere with each other both in the time domain and frequency domain. It is well known that the Iterative LMMSE detection greatly reduces the system computational complexity by departing the overall processing into many low-complexity distributed calculations that can be excuted in parallel. However, it is generally considered to be suboptimal and achieves relatively poor performance due to its suboptimal detector. In this paper, we firstly present the matching conditions and area theorems for the iterative detection of the MIMO-NOMA systems. Based on the matching conditions and area theorems, the achievable rate region of the Iterative LMMSE detection is analysed. Interestingly, we prove that by properly design the Iterative LMMSE detection, it can achieve \emph{(i)} the optimal capacity of symmetric MIMO-NOMA system, \emph{(ii)} the optimal sum capacity of asymmetric MIMO-NOMA system, \emph{(iii)} all the maximal extreme points in the capacity region of asymmetric MIMO-NOMA system, \emph{(iv)} the whole capacity region of two-user and three-user asymmetric MIMO-NOMA systems, in a distributed manner for all cases. Finally, a practical Iterative LMMSE detection design is also proposed for the general asymmetric MIMO-NOMA systems.
\end{abstract}


\IEEEpeerreviewmaketitle
\section{Introduction}

\emph{Non-Orthogonal Multiple Access} (NOMA) has attracted a lot of attentions and been recognized one of the key radio access technologies for the fifth generation (5G) mobile networks \cite{5GWhitepaper, METIS, Saito2013, Al-Imari2014, Ding2014, Ding2015, Ding20161, Ding20162, Kim2015, Chen2015}. The key concept behind NOMA is that all the users are allowed to be superimposed at the receiver in the same time/code/frequency domain to significantly increase the spectral efficiency and reduce latency in the 5G communictions systems \cite{ Saito2013, Al-Imari2014, Ding2014, Ding2015, Ding20161}. In addition, NOMA can be combined with the \emph{Multiuser Multiple-Input and Multiple-Output} (MU-MIMO)\cite{Argas2013, Rusek2013, biglieri2007, Marzetta2010}, which is a key technology for wireless communication standards like IEEE 802.11 (Wi-Fi), WiMAX and Long Term Evolution (4G), to further enhance the system performance \cite{Ding20162}. The MU-MIMO with NOMA is the well known MIMO-NOMA that is discussed in this paper.

Unlike the \emph{Orthogonal Multiple Access} (OMA) systems, \emph{e.g.} the \emph{Time Division Multiple Access }(TDMA) and \emph{Orthogonal Frequency Multiple Access} (OFMA) \emph{ect.}, the signal processing in the NOMA systems will cost higher complexity and higher energy consumption at the \emph{base station} (BS)\cite{Rusek2013,biglieri2007}. Low-complexity uplink detection for MIMO-NOMA is a challenging problem due to the non-orthogonal interference between the users \cite{Rusek2013, Al-Imari2014, Kim2015, Chen2015}, especially when the number of users and the number of antennas are large. The optimal \emph{multiuser detector} (MUD) for the MIMO-NOMA, such as the \emph{maximum a posterior probability} (MAP) detector or \emph{maximum likelihood} (ML) detector, was proven to be an NP-hard and non-deterministic polynomial-time complete (NP-complete) problem \cite{Micciancio2001,verdu1984_1}. Furthermore, the complexity of optimal MUD grows exponentially with the number of users or the number of antennas at the BS,  and polynomially with the size of the signal constellation \cite{verdu1986,verdu1984_2,verdu1987}. Many low-complexity linear detections such as \emph{Matched Filter} (MF), \emph{Zero-Forcing} (ZF) receiver \cite{tse2005}, \emph{Minimum Mean Square Error} (MMSE) detector and \emph{Message Passing Detector} (MPD) \cite{Loeliger2004, Loeliger2002, Loeliger2006} are proposed for the practical systems. In addition, some iterative methods such as \emph{Jacobi method}, \emph{Richardson method}\cite{Axelsson1994, Bertsekas1989, Gao2014}, \emph{Belief Propagation} (BP) method, and iterative MPD \cite{ Lei2015, andrea2005, Roy2001, yoon2014} are put forward to further reduce the computational complexity by avoiding the unfavorable matrix inversion in the linear detections. Although these detectors are attractive from the complexity view point, they achieve relatively poor performance and are considered to be sub-optimal MUDs for the MIMO-NOMA systems.

\emph{Successive Interference Cancellation} (SIC) based on the low-complexity detections is one of the key technology to improve the system performance. It is well known that for the \emph{multiuser access channel} (MAC), the SIC is an optimal strategy and can achieve the whole multiuser capacity region with time-sharing technology \cite{Cover2006, Gamal2012}. 
In \cite{verdu1998,Golden1999}, it is showed that the ZF-SIC detector is sub-optimal and can not achieve the full diversity of the MIMO system. In addition, the MMSE-SIC detector \cite{Wang1999, Choi2000,Studer2011} has been proposed to achieve optimal performance \cite{tse2005}. However, the SIC decoding still has some disadvantages when applying to the practical MIMO-NOMA systems, because it needs to preset the decoding order for all the users and assumes that all the previous users' messages are recovered correctly before we decode the next user's message. Furthermore, the decoding order changes with the different channel state and different \emph{Quality of Service} (QoS). These greatly increase the time delay, introduce error propagation during the decoding process, and require additional spectrum resource for the users and base station to achieve the preseted decoding order. To achieve the whole capacity region of the MIMO-NOMA systems, time-sharing should be used, which needs cooperation between the users. In addition, the complexity of SIC decoding is also too high to apply to the real system when the number of antennas at the BS is large \cite{Rusek2013,tse2005,verdu1998}. 

The efficient iterative detection that exchanges soft information of the low-complexity detector with the user decoders is widely used for the practical MIMO-NOMA systems \cite{andrea2005, Gao2014, Lei2015}. This is a fundamental technology for the non-orthogonal MAC like the \emph{Code Division Multiple Access} (CDMA) systems \cite{tse2005,verdu1998} and the \emph{Interleave Division Multiple Access} (IDMA) systems \cite{Ping2003_1,Ping2004,Ping2004_2}. Various iterative detectors, such as Iterative \emph{Linear MMSE} (LMMSE) detector, iterative BP detector and iterative MPD, are proposed to achieve a good performance \cite{Guo2008, wu2014, Sanderovich2005, Caire2004, Yuan2014}. The iterative detection is a low-complexity parallel joint decoding method, in which the time delay and error propagation can be greatly reduced. In addition, no user cooperation and additional spectrum resource will be wasted on the preseted decoding order in this method, and the complexity becomes much lower at the same time as the overall receiver is departed into many parallel distributed processors. However, in general, the joint iterative detection structure cannot achieve the perfect performance and is considered to be sub-optimal \cite{verdu1998}. Therefore, the achievable rate region analysis of the MIMO-NOMA systems with iterative detection is an intriguing problem.

The \emph{Extrinsic Information Transfer} (EXIT) \cite{Ashikhmin2004,Brink2001}, \emph{MSE-based Transfer Chart} (MSTC) \cite{Guo2005,Bhattad2007,Yuan2014}, area theorem and matching theorem \cite{Ashikhmin2004, Brink2001, Bhattad2007, Yuan2014, Guo2005} are the main analysis methods of the system achievable rate or the system \emph{Bit Error Rate} (BER) performance. It is proved that a well-designed single-code with linear precoding and Iterative LMMSE detection achieves the capacity of the MIMO systems \cite{Yuan2014}. However, this results is based on the \emph{point-to-point} (P2P) MIMO systes. For the multi-user MIMO-NOMA systems, this problem becomes much more complicated, because the whole achievable rate region that contains the achievable rates of all the users remains unknown. In addition, in the multi-user MIMO-NOMA systems, the users interfere with each other and their decoding processes interact each other based on the detector, which results in a much more complicated MST functions and area theorems. As a result, it is more difficult to analyse the achievable rates of all users for the MIMO-NOMA systems.

In this paper, we consider the uplink MIMO-NOMA communication network with a low-complexity Iterative LMMSE detection, where the multiple users communicate with the BS using the same time, frequency, and spreading code resources. The achievable rate region analysis of the Iterative LMMSE detection is provided, which shows that Iterative LMMSE can be rate region optimal for the MIMO-NOMA systems if properly designed. The contributions of this paper are summarized as follows.

\noindent
\hangafter=1
\setlength{\hangindent}{2em} a) For the MIMO-NOMA systems, matching conditions of the iterative detection are proposed. Based on these matching conditions, area theorems of MIMO-NOMA systems with iterative detection are proposed.

 \noindent
\hangafter=1
\setlength{\hangindent}{2em} b) With the proposed matching conditions and area theorems, the design and achievable rate analysis of Iterative LMMSE detection for the MIMO-NOMA systems are provided.

 \noindent
\hangafter=1
\setlength{\hangindent}{2em} c) We prove that the properly designed Iterative LMMSE detection \emph{(i)} is capacity achieving for symmetric MIMO-NOMA systems, \emph{(ii)} is sum capacity achieving for the asymmetric MIMO-NOMA systems, \emph{(iii)} achieves all the maximal extreme points in the capacity region of asymmetric MIMO-NOMA system, and \emph{(iv)} achieves the whole capacity region of two-user and three-user asymmetric MIMO-NOMA systems.

 \noindent
\hangafter=1
\setlength{\hangindent}{2em} d) For the general asymmetric MIMO-NOMA systems, a numerical algorithm for the practical Iterative LMMSE detection design is provided.

This paper is organized as follows. In Section II, the MIMO-NOMA system model and iterative detection are introduced. The matching conditions and area theorems for the MIMO-NOMA systems are elaborated in Section III. Section IV provides the achievable rate region analysis for the MIMO-NOMA systems with Iterative LMMSE detection. Some special cases are shown in Section V, and we end with conclusions in Section VI.

\section{System Model and Iterative Detection}
In this section, the MIMO-NOMA system model and some preliminaries about the iterative detection for the MIMO-NOMA systems are introduced.
\subsection{System Model}
\begin{figure}[ht]
  \centering
  \includegraphics[width=15cm]{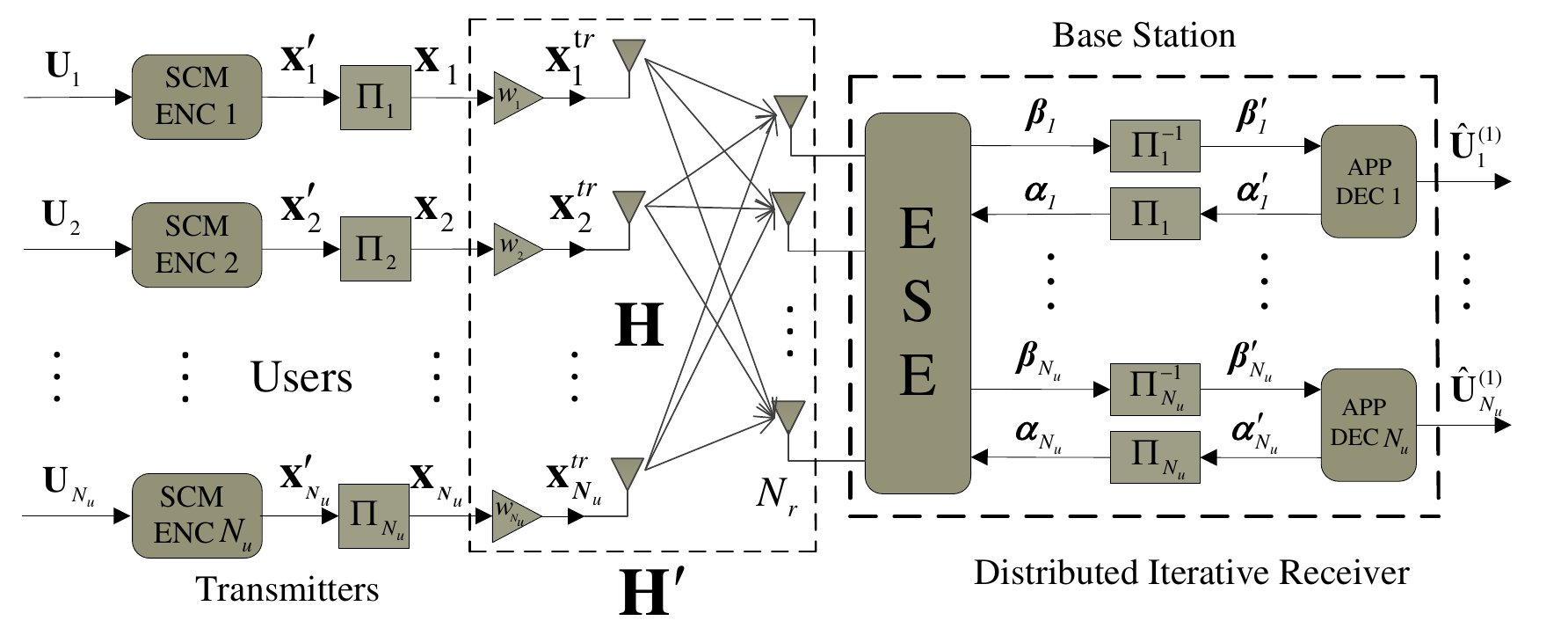}\\
  \caption{Block diagram of the MIMO-NOMA system. SCM ENC is the Superposition Coded Modulation Encoder and APP DEC is the A Posteriori Probability Decoder. $\Pi_i$ denotes the Interleaver and $\Pi_i^{-1}$ denotes the De-Interleaver. ESE represents the Elementary Signal Estimator. $\mathbf{H}'$ contains the small-scale fading channel $\mathbf{H}$, and $w_i^2$, $i\in\mathcal{N}_u $ denote the power constraint parameters or the large scale fading of the users.}\label{f2}
\end{figure}
Consider a uplink MU-MIMO system as showed in Fig \ref{f2}. In this system, $N_{u}$ autonomous single-antenna terminals simultaneously communicate with an array of $N_{r}$ antennas of the BS \cite{Marzetta2010,Rusek2013} at the same time and under the same frequency. Here, $N_{u}$ and $N_{r}$ can be any finite positive integers. Unlike the TDMA systems that the user are orthogonal in the time domain or the OFMA systems that the user are orthogonal in the frequency domain, all the users interfere with each other at the receiver and are non-orthogonal both in the time domain and frequency domain in the NOMA systems. This is the reason why we call it MIMO-NOMA. Then, the $N_{r}\times1$ received signal vector ${ \mathbf{y}(t)}$ at time $t$ at the BS can be represented as
\begin{equation}\label{e1}
{ \mathbf{y}(t)}= \mathbf{H}{ \mathbf{x}^{tr}(t)} + \mathbf{n}(t),\quad t\in \mathcal{N},
\end{equation}
where $\mathcal{N}=\{1,\cdots,N\}$, $\mathbf{H}$ is a $N_{r} \times N_{u}$ channel matrix, $\mathbf{n}(t)\sim\mathcal{CN}^{{N_{r}}}(0,\sigma_n^2)$ is the $N_{r} \times 1$ independent additive white Gaussian noise (AWGN) vector at time $t$, $\mathbf{x}^{tr}(t)=[x^{tr}_1(t),\cdots,x^{tr}_{N_u}(t)]^T$ is the message vector sent from $N_{u}$ users and  $\mathbf{y}(t)$ is the message vector got at the $N_{r}$ receive antennas at time $t$. In this paper, we consider the fading channels, which means $\mathbf{H}$ denotes the Rayleigh fading channel matrix whose entries are \emph{independently and identically distributed} (i.i.d.) with normal distribution $\mathcal{CN}(0,1)$, but it is fixed during the transmission.
The BS tries to estimate the sources from the received signal vector $\mathbf{y}(t)$, $t\in \mathcal{N}$. Note that the channel matrix $\mathbf{H}$ can be usually achieved by time-domain and/or frequency-domain training pilots. In this paper, we assume that the \emph{Channel State Information} (CSI) $\mathbf{H}$ is known at the BS .

\subsection{Transmitters}
As illustrated in Fig. \ref{f2}, at user $i$, an information sequence ${\bf{U}}_i $ is encoded by a channel code with rate $R_i$ into a $N$-length coded sequence ${\mathbf{x}}'_{i}$, $\mathop{i}\in \mathcal{N}_{u},\; \mathcal{N}_{u}= \left\{ {{{1,2,}} \cdots {{,N_{u}}}} \right\}$ and then interleaved by an {$N$-length} independent random interleaver $\Pi_{i}$ and get $\mathbf{x}_{i}=[x_{i,1},x_{i,2},\cdots,x_{i,N}]^T$. The interleavers here is used for improving the system performance by enhancing the randomness of the messages or the channel noise and avoiding the short cycles in the system factor graph \cite{Ping2003_1, Ping2004,Andrews2007}. We assume that each $x_{i,t}$ is randomly and uniformly taken over the points in a discrete signaling constellation $\mathcal{S}=\{s_1,s_2,\cdots,s_{|\mathcal{S}|}\}$. This assumption does not lose any generality since signal shaping (e.g., to approach Gaussian signaling) can be realized by properly designing the constellation points of $\mathcal{S}$. After that, the $\mathbf{x}_{i}$ is scaled with $w_i$, which denotes the power constraint or the large-scale fading coefficient of each user, and we then get the transmitting $\mathbf{x}^{tr}_{i}$, $\mathop{i}\in \mathcal{N}_{u}$. Let $K_{\mathbf{x}}$ be power constraint diagonal matrix whose diagonal elements are $w^2_i, \mathop{i}\in \mathcal{N}_{u}$. Therefore, the system model can be rewritten as
\begin{eqnarray}\label{e2}
{ \mathbf{y}_t}&=&\mathbf{H}{ K_{\mathbf{x}}^{1/2}\mathbf{x}(t)} + \mathbf{n}(t)\nonumber\\
&=& \mathbf{H}'{\mathbf{x}(t)} + \mathbf{n}(t),\quad t\in \mathcal{N},
\end{eqnarray}
where $K_{\mathbf{x}}^{1/2}=diag\{w_1,\cdots,w_{N_u}\}$, $\mathbf{H}'=\mathbf{H} K_{\mathbf{x}}^{1/2}$ is an equivalent channel matrix.

\subsection{Iterative Receiver}

At the base station, the received signals $\mathbf{Y}=[\mathbf{y}_1,\cdots,\mathbf{y}_N]$ and message $\bm{\alpha}_i$ from the decoder are sent to a low-complexity elementary signal estimator (ESE) to estimate the extrinsic message $\bm{\beta}_i$, which is then deinterleaved with $\Pi_i^{-1}$ into $\bm{\beta}'_i$. The corresponding single-user decoder employs $\bm{\beta}'_i$ as the prior message to calculate the extrinsic message $\bm{\alpha}'_i$. Similarly, this extrinsic message is interleaved by $\Pi_i$ to obtain the prior information $\bm{\alpha}_i$ for the ESE. Repeat this process until the maximum number of iteration is achieved or the messages are recovered correctly. The messages $\bm{\alpha}_i$, $\bm{\alpha}'_i$, $\bm{\beta}_i$ and $\bm{\beta}'_i$, $i\in\mathcal{N}_u$, are defined as follows.

Let $\bm{\alpha}_i$ be an $N$-length message vector as
\begin{equation}\label{e3}
\bm{\alpha}_{i}=[\bm{\alpha}_{i,1}, \bm{\alpha}_{i,2},\cdots,\bm{\alpha}_{i,N} ]^T
\end{equation}
and $\bm{\alpha}_{i,t}$ is a vector containing the likelihood values of $x_{i,t}$, i.e.,
\begin{equation}\label{e4}
\bm{\alpha}_{i,t}=\{{\alpha}_{i,t}(1), {\alpha}_{i,t}(2),\cdots,{\alpha}_{i,t}(|\mathcal{S}|) \},
\end{equation}
where ${\alpha}_{i,t}(k)$ denotes the prior likelihood of $x_{i,t}=s_k\in\mathcal{S}$ for the detector, and
\begin{equation}\label{e5}
\sum\limits_{k = 1}^{|S|} {{\alpha _{i,t}}(k) = } 1.
\end{equation}
Similar definitions applies to $\bm{\beta}'_i$, $\bm{\beta}_i$ and $\bm{\alpha}'_i$. The auxiliary random variables $a_{i,t}$ and $b'_{i,t}$ denote the information carried by $\bm{\alpha}_{i,t}$ and $\bm{\beta}'_{i,t}$ respectively, which are defined as
\begin{equation}\label{ea1}
p(a_{i,t}|x_i(t)=s_k)=\alpha_{i,t}(k)\;\;\;\; and \;\;\;\; p(b'_{i,t}|x'_i(t)=s_k)=\beta'_{i,t}(k).
\end{equation}
Let $\mathbf{a}_i=[a_{i,1},\cdots,a_{i,N_u}]$ and $\mathbf{b}'_i=[b'_{i,1},\cdots,b'_{i,N_u}]$, for $i\in\mathcal{N}$. To simplify to analysis of the iterative process, the following two assumptions are made for the ESE and decoders.

\emph{Assumption 1: For the estimator, each $x_i(t)$ is independently chosen from $\mathcal{S}$ with equal probability, i.e., $p(x_i(t)=s_k)=1/|\mathcal{S}|$, for any $i$, $k$ and $t$; the messages $\mathbf{a}_i$, $i\in {\mathcal{N}}_u$ are independent with each other, and the entries of $\mathbf{a}_i$ are i.i.d. given $\mathbf{x}_i$, i.e.,
 \begin{equation}\label{ea2}
p(\mathbf{a}_i|\mathbf{x}_i)=\prod\limits_{t=1}^{N}{p(a_{i,t}|x_{i,t})}
\end{equation}
and
  \begin{equation}\label{ea3}
p(a_{i,t_1}=s_l|{x}_{i,t_1}=s_k)=p(a_{i,t_2}=s_l|{x}_{i,t_2}=s_k)
\end{equation}
for any $i\in \mathcal{N}_u$, $s_k, s_l\in \mathcal{S}$ and $t_1, t_2\in \mathcal{N}$.}

\emph{Assumption 2: For the decoder, the messages $\mathbf{b}'_i$, $i\in {\mathcal{N}}_u$ are independent with each other, and the entries of $\mathbf{b}'_i$ are i.i.d. given $\mathbf{x}'_i$, i.e.,
 \begin{equation}\label{ea4}
p(\mathbf{b}'_i|\mathbf{x}'_i)=\prod\limits_{t=1}^{N}{p(b'_{i,t}|x'_{i,t})}
\end{equation}
and
  \begin{equation}\label{ea5}
p(b'_{i,t_1}=s_l|{x}'_{i,t_1}=s_k)=p(b'_{i,t_2}=s_l|{x}'_{i,t_2}=s_k)
\end{equation}
for any $i\in \mathcal{N}_u$, $s_k, s_l\in \mathcal{S}$ and $t_1, t_2\in \mathcal{N}$.}

Assumptions 1 and 2 decompose the overall process into the local processors such as the estimator and decoders, which simplifies the analysis of the iterative process. These assumptions are widely used in iterative decoding and turbo equalization algorithms. Actually, the messages $\bm{\beta}'_i$, $\bm{\beta}_i$, $\bm{\alpha}_i$ and $\bm{\alpha}'_i$ can be replaced by the auxiliary random variables $\mathbf{b}'_i$, $\mathbf{b}_i$, $\mathbf{a}_i$ and $\mathbf{a}'_i$ respectively.
\subsubsection{Element Signal Estimator (ESE)}
The extrinsic messages of the ESE is defined as
\begin{equation}\label{e6}
\beta_{i,t}(k)=p\left( x_{i,t}=s_k|\bm{\alpha}_{\sim i,t},\mathbf{y}_t \right), for \;\; i\in \mathcal{N}_u\;\; and\;\; t\in \mathcal{N},
\end{equation}
where $\bm{\alpha}_{\sim i,t}$ represents the vector obtained by deleting the $i$th entry of $[\bm{\alpha}_{1,t},\cdots,\bm{\alpha}_{N_u,t}]$.  Here, ``extrinsic" and ``elementary signal estimator" means that the output estimation for $x_{i,t}$, i.e., the $t$th transmission at user $i$, is based on the $t$th receiver $\mathbf{y}_t$ and the messages of the $t$th symbols from the decoders except user $i$.

For the MIMO-NOMA systems, it is too complicated to calculated (\ref{e6}) directly. Element signal estimator is an alternative low complexity estimator. For the ESE, with the input prior messages $\bm{\alpha}_{i,t}$, we can calculate the prior mean and variance of $x_i(t)$, respectively, by
\begin{equation}\label{e7}
{{\bar x}_{i,t}} = E\left[{x_{i,t}}|{\bm{\alpha}_{i,t}}\right] = \sum\limits_{k = 1}^{|S|} {{\alpha _{i,t}}(k){s_k}},
\end{equation}
and
\begin{equation}\label{e8}
v_i = E\left[|{x_{i,t}}-\bar{x}_{i,t}|^2\right]=\sum\limits_{k = 1}^{|S|}{\alpha_{i,t}(k)|s_k-\bar{x}_{i,t}|^2},
\end{equation}
for any index $t$. From Assumption 1, $v_i$ is invariant with respect to the index $t$. Then, the input messages $\bm{\alpha}_{i,t}$ can be replaced by the ${{\bar {{x}}}_{i,t}}$ and ${v}_i$ if the messages are Gaussian distributed.

Let $\hat{x}_{i,t}$ be the total estimation of ${x}_{i,t}$ at the detector, and $v_{\hat{x}_i}$ be the corresponding estimated deviation. The extrinsic mean and variance for $x_{i,t}$ (denoted by $u_i$ and $b_{i,t}$) by excluding the contribution of the prior message $\bm{\alpha}_i$ according to the message combining rule \cite{Loeliger2004,Loeliger2002,Loeliger2006} as
\begin{equation}\label{e9}
u_i^{-1} = v_{\hat{x}_i}^{-1}-v_i^{-1}
\end{equation}
and
\begin{equation}\label{e10}
\frac{b_{i,t}}{u_i}=\frac{\hat{x}_{i,t}}{v_{\hat{x}_i}}-\frac{\bar{x}_{i,t}}{v_{i}}.
\end{equation}
Finally, the extrinsic message of the estimator can be given by
\begin{equation}\label{e11}
\beta_{i,t}(k)=\frac{p(b_{i,t}|x_{i,t}=s_k)}{\sum\limits_{k = 1}^{|S|}p(b_{i,t}|x_{i,t}=s_k) }.
\end{equation}

The overall complexity has been reduced by the local processing of the ESE and distributed decoders. This is very important for the application for the practical systems. We will show that not only the complexity of the iterative receiver are greatly reduced, but also it is sum capacity achieving if the ESE and channel codes of each user are properly designed.

\subsubsection{\emph{A Posteriori Probability} (APP) Decoding}
The decoder employees the APP decoding based on the input message $\bm{\beta}_i$, $i\in \mathcal{N}_u$. The output extrinsic message of the decoder is defined as
\begin{equation}\label{e16}
\alpha'_{i,t}(k)=p(x_i(t)=s_k|\bm{\beta}'_{\sim i}),
\end{equation}
where $i\in \mathcal{N}_u$, $t\in \mathcal{N}$ and $k\in \mathcal{S}$.

\emph{Assumption 3: The local decoder is APP decoding.}

Although computational complexity of the APP decoding is too high to apply in practical systems, low-complexity message-passing algorithms can be used to achieve near-optimal performance \cite{Richardson2001}. Assumption 3 is introduced to simplify our analysis. As the message passing decoding is well-studied in the literature, and thus the details are omitted here.

\subsection{Capacity region of the MIMO-NOMA systems}
 Let $\mathbf{Y}$ denote the received random variable vector at the BS, and $\mathbf{X}$ represent the received random variable vector from the users. Assume $S\subseteq \mathcal{N}_u$, $S^c\subseteq \mathcal{N}_u/S$ and $S\cup S^c=\mathcal{N}_u$, the partial channel matrix is denoted as $\mathbf{H}'_S=[\mathbf{h}'_i]_{N_r\times |S|}$, where $\mathbf{h}'_i$ is the $i$th column of $\mathbf{H}'$ and $i\in \mathcal{S}$. Similar definition applies to the partial random vector $\mathbf{X}_S$ and the partial matrix $K_{\mathbf{x}_S}$. The rate of user $i$ is $R_i$ and $R_S=\sum\limits_{i\in S}{R_i}$ represents the sum rate of the users in set $S$. Then, for our MIMO-NOMA systems (\ref{e2}), the capacity region $\mathbf{\mathcal{R}}_\mathcal{S}$ is given as follows \cite{Cover2006, Gamal2012}.
\begin{eqnarray}\label{e17}
{R_S} &\le& I(\mathbf{Y};{\mathbf{X}_S}|{\mathbf{X}_{{S^c}}})\nonumber\\
& =& \log\det\left(\mathbf{I}_{|\mathcal{S}|} + \frac{1}{\sigma_n^2}\mathbf{H}'^H_S\mathbf{H}'_S\right),
 \;\mathrm{ for }\; \forall S \subseteq \mathcal{N}_u\; \mathrm{and} \; S^c\subseteq \mathcal{N}_u/S.
\end{eqnarray}
Therefore, the sum rate of the MU-MIMO systems is denoted as
\begin{eqnarray}\label{e18}
R_{sum}&=& R_{\mathcal{N}_u} \le I(\mathbf{Y};{\mathbf{X}_{\mathcal{N}_u}})\nonumber\\
& =& \log\det\left(\mathbf{I}_{N_u} + \frac{1}{\sigma_n^2}\mathbf{H}'^H\mathbf{H}'\right).
\end{eqnarray}

{\emph{Capacity Region Domination Lemma}}: \emph{\cite{Han1979} All point in the capacity region $\mathbf{\mathcal{R}}_\mathcal{S}$ is dominated by a convex combination of the following $(N_u!)$ maximal extreme points.}
\begin{equation}\label{dominate}
\left\{ \begin{array}{l}
{R_{{k_1}}}  = I(\mathbf{Y};{\mathbf{X}_{{\mathcal{S}_1}}}) = \log\det\left(\mathbf{I}_{N_u} + \frac{1}{\sigma_n^2}\mathbf{H}'^H\mathbf{H}'\right) - \log\det\left(\mathbf{I}_{|\mathcal{S}_{1}^c|} + \frac{1}{\sigma_n^2} \mathbf{H}_{\mathcal{S}_{1}^c}'^H\mathbf{H}_{\mathcal{S}^c_{1}}'\right), \\
    \qquad\qquad\vdots \\
{R_{{k_{N_u-1}}}}  \!\!= I(\mathbf{Y};{\mathbf{X}_{{k_{N_u-1}}}}|{\mathbf{X}_{{{\mathcal{S}_{N_u-2}}}}}) = \log\det\left(\mathbf{I}_{|\mathcal{S}_{N_u-2}^c|} + \frac{1}{\sigma_n^2}\mathbf{H}_{\mathcal{S}^c_{N_u-2}}'^H \mathbf{H}_{\mathcal{S}^c_{N_u-2}}'\right)\! -\! \log \left( \!{1 + \frac{1}{{\sigma _n^2}}{\mathbf{h}'}_{{k_{N_u}}}^H{{\mathbf{h}'}_{{k_{N_u}}}}}\! \right), \\
{R_{{k_{{N_u}}}}}  = I(\mathbf{Y};{\mathbf{X}_{{k_{{N_u}}}}}|{\mathbf{X}_{{{\mathcal{S}_{N_u-1}}}}}) = \log\det\left(\mathbf{I}_{|\mathcal{S}_{N_u-1}^c|} + \frac{1}{\sigma_n^2} \mathbf{H}_{\mathcal{S}^c_{N_u-1}}'^H \mathbf{H}_{\mathcal{S}^c_{N_u-1}}'\right) = \log \left( {1 + \frac{1}{{\sigma _n^2}}{\mathbf{h}'}_{{k_{N_u}}}^H{{\mathbf{h}'}_{{k_{N_u}}}}} \right),\\
\end{array} \right.
\end{equation}
\emph{where $(k_1,
\cdots,k_{N_u})$ is a permutation of $(1,2,\cdots,N_u)$, $\mathcal{S}_i=\{k_1,\cdots,k_i\}$ and $\mathcal{S}_i^c=\mathcal{N}_u/\{k_1,\cdots,k_i\}$ for $i=1,\dots,N_u-1$.}

\section{Matching Conditions and Area Theorems for MIMO-NOMA Systems}
In this section, we proposed the matching conditions and area properties of the iterative MIMO-NOMA systems from their \emph{SINR-Variance} transfer functions. In \cite{Guo2005, Bhattad2007,Yuan2014}, the \emph{I-MMSE} theorem and the area theorems for the P2P communication systems are proposed for its achievable rate analysis. In this section, these results are generalized to the MIMO-NOMA systems, which is much more complicated than the P2P case as all the users' singals interfere with each other and their decoding processes interact each other based on the detector. We will see that the achievable rate of each user in the iterative MIMO-NOMA systems can be derived based on these useful theorems and techniques.


\subsection{Characterization of ESE}
We first define the \emph{SINR-Variance} transfer function of user $i$ of element signal estimator as
\begin{equation}
\phi_i(\mathbf{v}_{\bar{\mathbf{x}}})=v_{\hat{x}_i}^{-1}-v_i^{-1},\;\; \mathrm{for}\;\; i\in \mathcal{N}_u,
\end{equation}
where $\mathbf{v}_{\bar{\mathbf{x}}}=[v_1,\cdots,v_{N_u}]$, $v_{\hat{x}_i}$ is the $i$th diagonal element of covariance matrix $\mathbf{V} _{{{\hat {\mathbf{x}}}}}$, and $v_i$ is the input variance of user $i$, i.e., the $i$th diagonal element of $\mathbf{V} _{{{ \bar{\mathbf{x}}}}}$. Actually, $\phi_i(\mathbf{v})$ denotes the output extrinsic \emph{SINR} of user $i$ at the estimator. Similarly, the total MSE of user $i$ at the estimator is
\begin{equation}\label{e19}
\mathrm{mmse}_{tot,i}^{ese}(\mathbf{v}_{\bar{\mathbf{x}}})=v_{\hat{x}_i}.
\end{equation}

It should be noted that the variance $v_i$ varies from $0$ to $1$, because the signal power is normalized to 1. From (\ref{e18}), the output estimation of user $i$ depends on the input variances of all the users. Thus, the \emph{SINR-Variance} transfer functions of all users interact with each other. In addition, $\phi_i (\mathbf{v}_ {\bar{\mathbf{x}}})$ is monotonically decreasing in $\mathbf{v}$, which means the lower input variances of the users, the higher the output \emph{SINR} of the estimator. The next Gaussian assumption is used to simplify the system analysis, which is a common assumption in many works \cite{Kay1993,Poor1997}.

\emph{Assumption 4: Let $\bm{\rho}=[\rho_1,\cdots,\rho_{N_u}]$, $\bm{\phi}(\mathbf{v}_{\bar{\mathbf{x}}})= \left[{\phi_1}(\mathbf{v}_{\bar{\mathbf{x}}}),\cdots, {\phi_{N_u}}(\mathbf{v}_{\bar{\mathbf{x}}}) \right]$. The outputs of the estimator can be approximated as the observations from AWGN channels and the related \emph{SINR} is denoted by $\bm{\rho}=\bm{\phi}(\mathbf{v}_{\bar{\mathbf{x}}})$. }

\subsection{Characterization of APP Decoder}
We now consider the \emph{SINR-Variance} transfer functions of the decoders. From Assumption 4, the input of the each decoder $\mathbf{b}'_i$ are equivalent as the independent observations over an AWGN channel with $SNR_i=\rho_i$. As the LMMSE estimator only depends on the variance of input messages, we thus use the \emph{SINR-Variance} transfer functions to describe the decoders. The output extrinsic variance of APP decoder is defined as
\begin{equation}\label{e23}
 v'_i = v'_{i,t}= \mathrm{MMSE}\left(x'_{i,t}|\mathbf{b}'_{i,\sim t}\right)= E\left[ \left|x'_{i,t}-E[x'_{i,t}|\mathbf{b}'_{i,\sim t}] \right|^2\right],
\end{equation}
where the input messages $\bm{\beta}'_i$ are replaced by auxiliary random variable $\mathbf{b}'_i$, and the expectation is taken over the joint probability space of $\mathbf{x}'_i$ and $\mathbf{b}'_i$. The first equation is derived from the i.i.d. Assumption 2. Therefore, we define the \emph{SINR-Variance} transfer function of the decoders as
\begin{equation}\label{e24}
v_i=\psi_i(\rho_i)
\end{equation}
for any $i\in \mathcal{N}_u$, where $\psi_i$ can be different transfer function with different $i$. Let $\bm{\psi}(\bm{\rho})=[\psi_1(\rho_1),\cdots,\psi_{N_u}(\rho_{N_u})]$, and we have
\begin{equation}\label{e25}
\mathbf{v}_{\bar{\mathbf{x}}}=\bm{\psi}(\bm{\rho}).
\end{equation}

\subsection{SINR-Variance Transfer Chart and Matching Conditions}
The LMMSE estimator is described by $\bm{\rho}=\bm{\phi}(\mathbf{v}_{\bar{\mathbf{x}}})$, and the decoders can be described by $\mathbf{v}_{\bar{\mathbf{x}}}=\bm{\psi}(\bm{\rho})$. Therefore, the iterative detection performs iteration between the estimator and the decoders and can be tracked by $\bm{\rho}$ and $\mathbf{v}_{\bar{\mathbf{x}}}$. Let $\tau$ represent the $\tau$-th iteration. We then have
\begin{equation}\label{e27}
\bm{\rho}(\tau)=\bm{\phi}\left(\mathbf{v}_{\bar{\mathbf{x}}}(\tau-1)\right) \;\mathrm{and}\;\mathbf{v}_{\bar{\mathbf{x}}}(\tau)=\bm{\psi}\left(\bm{\rho}(\tau)\right), \quad\tau=1,2,\cdots
\end{equation}
The iteration converges to a fix point $\mathbf{v}_{\bar{\mathbf{x}}}^*$ which satisfies
\begin{equation}\label{e28}
\bm{\phi}\left(\mathbf{v}_{\bar{\mathbf{x}}}^*\right) =\bm{\psi}^{-1}\left(\mathbf{v}_{\bar{\mathbf{x}}}^*\right)\; \mathrm{and} \; \bm{\phi}\left(\mathbf{v}_{\bar{\mathbf{x}}}\right) >\bm{\psi}^{-1}\left(\mathbf{v}_{\bar{\mathbf{x}}}\right),\quad \mathrm{for} \; \mathbf{v}_{\bar{\mathbf{x}}}^*  < \mathbf{v}_{\bar{\mathbf{x}}} \leq \mathbf{1},
\end{equation}
where $\bm{\psi}^{-1}(\cdot)$ denotes the inverse of $\bm{\psi}(\cdot)$ which exists since $\bm{\psi}(\cdot)$ is continuous and monotonic \cite{Guo2011}. It should be mentioned that the equations and inequalities for the vectors or matrixes in this paper correspond to the component-wise inequality. The inequality $\mathbf{v}_{\bar{\mathbf{x}}} \leq \mathbf{1}$ comes from the normalized signal powers of $\mathbf{x}(t)$, $t\in \mathcal{N}$. If $\mathbf{v}_{\bar{\mathbf{x}}}^*=\mathbf{0}$, then all the transmissions can be recovered correctly, which means that $\bm{\phi}\left(\mathbf{v}_{\bar{\mathbf{x}}}\right) >\bm{\psi}^{-1}\left(\mathbf{v}_{\bar{\mathbf{x}}}\right)$ for any available $\mathbf{v}_{\bar{\mathbf{x}}}$, i.e., the inverse transfer function of the decoder $\bm{\psi}^{-1}\left(\mathbf{v}_{\bar{\mathbf{x}}}\right)$ lies below the transfer function of the estimator $\bm{\phi}\left(\mathbf{v}_{\bar{\mathbf{x}}}\right)$.

The estimator and decoders are matched if
\begin{equation}\label{e29}
\bm{\phi}\left(\mathbf{v}_{\bar{\mathbf{x}}}\right) =\bm{\psi}^{-1}\left(\mathbf{v}_{\bar{\mathbf{x}}}\right),\quad \mathrm{for} \;\; \mathbf{0}  < \mathbf{v}_{\bar{\mathbf{x}}} \leq \mathbf{1}.
\end{equation}
It means that $\phi_i(\mathbf{v}_{\bar{\mathbf{x}}})=\psi_i^{-1}(v_i)$ for any $i\in \mathcal{N}_u$. The matched transfer function not only maximizes the rate of the code, but also ensures the transmitting signals can be perfectly recovered. Note that: $\phi_i(\mathbf{1})>0$ as the estimator always use the information from the channel; and $\phi_i(\mathbf{0})>1$ as the estimator cannot remove the uncertainty introduced by the channel noise. Therefore, we have the following proposition.

\emph{Proposition 1}: \emph{For any $i\in \mathcal{N}_u$, the matching conditions of the iterative MIMO-NOMA systems can be rewritten as}
\begin{eqnarray}
\psi_i(\rho_i)&=&\phi_i^{-1}(\phi_i(\mathbf{1}))=1, \;\;\mathrm{for}\;\; 0\leq\rho_i<\phi_i(\mathbf{1});\label{e30}\\
\psi_i(\rho_i)&=&\phi_i^{-1}(\rho_i),\;\;\mathrm{for}\;\; \phi_i(\mathbf{1})\leq\rho_i<\phi_i(\mathbf{0});\label{e31}\\
\psi_i(\rho_i)&=&0,\;\;\mathrm{for}\;\; \phi_i(\mathbf{0})\leq\rho_i<\infty.\label{e32}
\end{eqnarray}
Similarly, ${\phi}_i^{-1}(\cdot)$ denotes the inverse of ${\phi}_i(\cdot)$ corresponding to the variable $v_i$. These matching conditions are very important for the area theorems in the next section.

\subsection{Area Properties}
 Let $\mathrm{snr}_{ap,i}^{dec}$ denote the \emph{SNR} of the prior input messages in decoder $i$, $\mathrm{snr}_{ext,i}^{ese}$ be the \emph{SNR} of the output extrinsic messages of user $i$ at the estimator, $\mathrm{mmse}_{tot,i}^{ese}(\cdot)$ represent the total variance of the messages of user $i$ at the LMMSE estimator, and $\mathrm{mmse}_{tot,i}^{dec}(\cdot)$ indicate the total variance of the messages at decoder $i$.
In addition, $\mathbf{snr}_{ext,i}^{ese}=[\mathrm{snr}_{ext,1}^{ese}, \cdots,\mathrm{snr}_{est,N_u}^{ese}]$. Then, we present two area properties of the estimator and decoders as the following proposition, which will be used to derive the achievable rates of each user.

 \emph{Proposition 2}: \emph{The achievable rate $R_i$ of user $i$ and an upper bound of $R_i$ are given as}
 \begin{eqnarray}
{R_i} = \int\limits_0^\infty  {\mathrm{mmse}_{tot,i}^{dec}({{snr}}_{ap,i}^{dec})d} \mathrm{snr}_{ap,i}^{dec},\label{e33}\\
R_i^{\max } = \int\limits_0^\infty  {\mathrm{mmse}_{tot,i}^{ese}(\mathbf{snr}_{ext}^{ese})d} \mathrm{snr}_{ext,i}^{ese},\label{e34}
\end{eqnarray}
\emph{and $R_i\leq R_i^{\mathrm{max}}$, $i\in \mathcal{N}_u$, where the equality holds if and only if the \emph{SINR-Variance} transfer functions of the element signal estimator and decoders for any user are matched with each other, i.e., the matching conditions (\ref{e29})$\sim$ (\ref{e32}) hold.}

In our MIMO-NOMA system model, from (\ref{e18}) and (\ref{e24}) and with the Gaussian assumptions, we have $\mathrm{snr}_{ap,i}^{dec}=\rho_i$, $\mathrm{snr}_{ext,i}^{ese}=\phi_i(\mathbf{v}_{\bar{\mathbf{x}}})$, $\mathrm{mmse}_{tot,i}^{dec}(\mathbf{{snr}}_{ap}^{dec})={{\left( {{\rho _i} + {\psi _i}{{({\rho _i})}^{ - 1}}} \right)}^{ - 1}}$ and $\mathrm{mmse}_{tot,i}^{ese}(\mathbf{snr}_{ext,i}^{ese})=v_{\hat{x}_i}(\mathbf{v}_{\bar{\mathbf{x}}})$.
Therefore, (\ref{e33}) and (\ref{e34}) can be rewritten as following.

\emph{Proposition 3}:\emph{ With the \emph{SINR-Variance} transfer functions $\bm{\rho}=\bm{\phi}(\mathbf{v}_{\bar{\mathbf{x}}})$ and $\mathbf{v}_{\bar{\mathbf{x}}}=\bm{\psi}(\bm{\rho})$,} the achievable rate $R_i$ of user $i$ and an upper bound of $R_i$ are
 \begin{eqnarray}
{R_i} = \int\limits_0^\infty  {{{\left( {{\rho _i} + {\psi _i}{{({\rho _i})}^{ - 1}}} \right)}^{ - 1}}d{\rho _i}} ,\label{e35}\\
R_i^{\max } = \int\limits_0^\infty  v_{\hat{x}_i}(\mathbf{v}_{\bar{\mathbf{x}}}) d \phi_i(\mathbf{v}_{\bar{\mathbf{x}}}).\label{e36}
\end{eqnarray}
\emph{and $R_i\leq R_i^{\mathrm{max}}$, $i\in \mathcal{N}_u$, where the equality holds if and only if the \emph{SINR-Variance} transfer functions of the element signal estimator and decoders for any user are matched with each other£¬ i.e., the matching conditions (\ref{e29})$\sim$ (\ref{e32}) hold.}

Thus, the achievable rates can be calculated directly by (\ref{e36}) or be calculated directly by (\ref{e35}) together with (\ref{e29}) and the matching conditions (\ref{e30})$\sim$ (\ref{e32}). From (\ref{e18}), (\ref{e24}), (\ref{e29}) and (\ref{e36}), we can see that all the users' transfer functions interact with each other at the estimator since every output of the estimator depends on the variances of the input messages from all the decoders. In addition, all the users' transfer functions are unknown and need to be properly designed. Thus, it is very hard to calculate the achievable rates directly with (\ref{e35}) and (\ref{e36}). However, for some special cases or with some additional constraints on the users' transfer functions, as we will see in the next section, the achievable rate of each user can then be calculated.

\section{Achievable Rate Region Analysis of Iterative LMMSE Detection}
In this section, based on the proposed matching conditions and area properties, the achievable rate of each user is given for the Iterative MIMO-NOMA systems. Here, we use the LMMSE estimator as the ESE estimator and the Superposition Coded Modulation (SCM) codes as the channel codes. As it is very difficult to analyse the achievable rate region for the general case, we first consider the symmetric MIMO-NOMA systems, where the users have the same rate and the same power. We prove that the Iterative LMMSE detection achieves the capacity of symmetric MIMO-NOMA system. This means that for the symmetric MIMO-NOMA systems, the Iterative LMMSE detection is optimal if the channel codes are properly designed. For the general asymmetric MIMO-NOMA systems, we also prove that the sum rate of the Iterative LMMSE detection achieves the sum capacity.

\subsection{LMMSE ESE}
LMMSE is an alternative low complexity ESE. In general, LMMSE detection is suboptimal for the multi-user large-scale MIMO-NOMA system with the discrete (or digital) sources \cite{Rusek2013}. However, for the gaussian sources, LMMSE detection is an optimal linear detector under MSE measure because it minimizes the MSE between sources and estimation \cite{verdu1998}. Let $\bar{\mathbf{x}}(t)=[{x_{1,t}},\cdots,x_{N_u,t}]$ and $\mathbf{V}_{\bar{\mathbf{x}}(t)}\!=\!\mathbf{V}_{\bar{\mathbf{x}}}= \mathrm{diag}(v_1,v_2,\cdots,v_{N_u})$. The LMMSE detector \cite{tse2005} is given by
\begin{eqnarray}\label{GMP2}
{{\hat {\mathbf{x}}}(t)} &=& \mathbf{V}^{-1}_{\hat {\mathbf{x}}}\left[\mathbf{V}_{\bar{\mathbf{x}}}^{-1}\bar{\mathbf{x}}(t)+ \sigma^{-2}_n\mathbf{H}'^H\mathbf{y}_t  \right]\nonumber\\
&=& \left(\sigma_n^{-2}\mathbf{H}'^H\mathbf{H}'+ \mathbf{V}_{\bar{\mathbf{x}}}^{-1}\right)^{-1}\left[\mathbf{V}_{\bar{\mathbf{x}}}^{-1}\bar{\mathbf{x}}(t)+ \sigma_n^{-2}\mathbf{H}'^H\mathbf{y}_t  \right],
\end{eqnarray}
where $\mathbf{V} _{{{\hat {\mathbf{x}}}}} = (\sigma _{{{n}}}^{- 2}\mathbf{H}'^H\mathbf{H}'+\mathbf{V} _{{{ \bar{\mathbf{x}}}}}^{-1})^{-1}$, which denotes the deviation of the estimation to the sources.

{\textit{Matrix Inversion Lemma:}} \emph{Let $\mathbf{A}$ and $\mathbf{D}$ be positive definite matrixes. For matrixes $\mathbf{B}$ and $\mathbf{C}$ with proper size, we have}
\begin{equation}\label{EQ1}
(\mathbf{A}\! +\! \mathbf{B}{\mathbf{D}^{ - 1}}\mathbf{C})^{-1}\!\! =\! {\mathbf{A}^{ - 1}} \!-\! {\mathbf{A}^{ - 1}}\mathbf{B}{(\mathbf{D}\! +\! \mathbf{C}{\mathbf{A}^{ - 1}}\mathbf{B})^{ - 1}}\mathbf{C}{\mathbf{A}^{ - 1}}.
\end{equation}

With the Matrix Inversion Lemma, the LMMSE estimator (\ref{GMP2}) can be rewritten as
\begin{eqnarray}\label{e20}
{{\hat {\mathbf{x}}}(t)} &=& \left(\sigma_n^{-2}\mathbf{H}'^H\mathbf{H}'+ \mathbf{V}_{\bar{\mathbf{x}}}^{-1}\right)^{-1}\left[\mathbf{V}_{\bar{\mathbf{x}}}^{-1}\bar{\mathbf{x}}(t)+ \sigma_n^{-2}\mathbf{H}'^H\mathbf{y}_t  \right]\nonumber\\
&=&\left[\mathbf{V}_{\bar{\mathbf{x}}} - \mathbf{V}_{\bar{\mathbf{x}}} \mathbf{H}'^H\left(\sigma_n^2\mathbf{I}_{N_r}+ \mathbf{H}'V_{\bar{\mathbf{x}}}\mathbf{H}'^H \right)^{-1}\mathbf{H}'\mathbf{V}_{\bar{\mathbf{x}}} \right] \left[\mathbf{V}_{\bar{\mathbf{x}}}^{-1}\bar{\mathbf{x}}(t)+ \sigma_n^{-2}\mathbf{H}'^H\mathbf{y}_t  \right]\nonumber\\
&=& \bar{\mathbf{x}}(t) + V_{\bar{\mathbf{x}}}\mathbf{H}'^H\left(\sigma_n^2\mathbf{I}_{N_r}+ \mathbf{H}'V_{\bar{\mathbf{x}}}\mathbf{H}'^H \right)^{-1}\left(\mathbf{y}_t-\mathbf{H}'\bar{\mathbf{x}}(t)\right).
\end{eqnarray}
Therefore, with (\ref{e10}) and (\ref{e20}), we get
\begin{equation}\label{e21}
b_{i,t} = x_{i,t} + n^{*}_{i,t},
\end{equation}
and
\begin{equation}\label{e22}
n_{i,t}^{*}=  \frac{v_i}{v_{\hat{x}_i}\rho_i}v_{\bar{x}_i}^2{\mathbf{h}'_i}^H\left( \sigma^2_n\mathbf{I}_{N_r}+ \mathbf{H}'\mathbf{V}_{\bar{\mathbf{x}}}\mathbf{H}'^H \right)^{-1}\left[ \mathbf{H}'\left(\mathbf{x}_{\backslash i}(t)-\bar{\mathbf{x}}_{\backslash i}(t)\right) +\mathbf{n}(t)\right].
\end{equation}
where $\mathbf{x}_{\backslash i}(t)$ (or $\bar{\mathbf{x}}_{\backslash i}(t)$) denotes the vector whose $i$th entry of $\mathbf{x}(t)$ (or $\bar{\mathbf{x}}(t)$)  is set to zero. Thus, we rewrite the Assumption 4 as follows.

\emph{Assumption 5: The equivalent noise $n_{i,t}^{*}$ is independent of $x_{i,t}$ and is Gaussian distributed $n_{i,t}\sim \mathcal{CN}\left(0,1/\phi_i(\mathbf{v}_{\bar{\mathbf{x}}})\right)$, i.e., the output of the LMMSE estimator $\mathbf{b}_{t}$ is the observation from AWGN channel, i.e., $\mathbf{b}_{t}=\mathbf{x}(t)+\mathbf{n}_t^{*}$ with SNRs $\bm{\rho}=\bm{\phi}(\mathbf{v}_{\bar{\mathbf{x}}})$.}

\emph{Note:} Although the overall complexity has been reduced by the local ESE and the distributed decoding processing, the complexity of LMMSE estimator is still very high for the practical applications when the number of users and antennas very large. The high complexity is mainly introduced by the matrix inversion and matrix multiplications in (\ref{GMP2}). In the practical systems, a lower complexity Gaussian message passing iterative estimator can also be used as the ESE for the iterative receiver of the MIMO-NOMA systems, which was proved that it converges to the LMMSE detector \cite{Lei2015}.

\subsection{Superposition Coded Modulation APP Decoders}
In this subsection, we introduce an important property that is established in \cite{Yuan2014}, which builds the relationship between the rate of the Forward Error Correction (FEC) code and its transfer function $\psi_i(\rho_i)$. The result is based on the SCM \cite{Wachsmann1999,Gadkari1999} and the area theorems \cite{Guo2005,Bhattad2007}. For more details, please refer to \cite{Yuan2014}.

\emph{Property of SCM Codes}: \emph{Assume $\psi(\rho)$ satisfies the following regularity conditions:\\
(i) $\psi(0)=1$ and $\psi(\rho)\geq 0$, for $\rho\in[0,\infty);$\\
(ii) monotonically decreasing in $\rho\in[0,\infty)$;\\
(iii) continuous and differentiable in $[0,\infty)$ except for a countable set of values of $\rho$;\\
(iv) $\mathop {\lim }\limits_{{\rho } \to \infty } {\rho }{\psi }({\rho }) = 0$.\\
Let $\Gamma_n$ be an $n$-layer SCM code with \emph{SINR-variance} transfer function $\psi^n(\rho)$ and rate $R_n$. Then, there exist $\{\Gamma_n\!\}$ such that: (i) $\psi^n(\rho)\!\leq\!\psi(\rho)$, for any $\rho\!\geq\!0$ and any integer $n$; (ii) as $n\!\to\! \infty$,
\begin{equation}\label{e26}
{R_n} \to R \left(\psi(\rho)\right).
\end{equation}}

In (\ref{e26}), $R \left(\psi(\rho)\right)$ denotes the rate of the code whose transfer function is $\psi(\rho)$. It means that there exist such an $n$-layer SCM code $\Gamma_n$ whose transfer function can approach the function $\psi(\rho)$ that satisfies the conditions (i)$\sim$(iv) with arbitrary small error when $n$ is large enough.

\subsection{Capacity Achieving of Iterative LMMSE Detection for Symmetric Systems}
In the subsection, we consider the symmetric MIMO-NOMA system, that is the users have the same power and the same rate constraints, i.e., $K_{\mathbf{x}}=w^2I$ and $R_i=R_j$, for $ i,j\in\mathcal{N}_u$. In addition, we assume that the number of users $N_u$ and the number of antennas $N_r$ are very large (be hundreds or more), i.e., the large-scale MIMO-NOMA case. In this case, since all the users have the same conditions, we thus assume that all the users have the same transfer functions, which means $v_i=v$ and $\rho_i=\rho$, for any $ i\in\mathcal{N}_u$. The symmetric and large-scale systems assumptions are used to simplify the analysis of the achievable user rates. Actually, in the next subsection, the achievable rate analysis for the asymmetric systems with finite $N_r$ and $N_u$ shows that this result also works for the practical asymmetric systems.

Based on these assumptions, we have
 \begin{eqnarray}\label{e37}
v_{\hat{x}_i}(\mathbf{v}_{\bar{\mathbf{x}}}) &\approx& \frac{1}{N_u}\mathrm{mmse}_{tot}^{ese}(\mathbf{v}_{\bar{\mathbf{x}}})\nonumber\\
&=&\frac{1}{N_u} \mathrm{Tr}\{\mathbf{V} _{{{\hat {\mathbf{x}}}}}\} \nonumber\\
&=&\frac{1}{N_u} \mathrm{Tr}\{(\sigma _{{{n}}}^{- 2}\mathbf{H}'^H\mathbf{H}'+\mathbf{V} _{{{ \bar{\mathbf{x}}}}}^{-1})^{-1}\}\nonumber\\
&=&\frac{1}{N_u} \mathrm{Tr}\{\left(\sigma _{{{n}}}^{- 2}w^2\mathbf{H}^H\mathbf{H}+v^{-1}\mathbf{I}_{N_u}\right)^{-1}\}\nonumber\\
&=&v_{\hat{x}}(v),
\end{eqnarray}
and
\begin{eqnarray}\label{e38}
\phi_i(\mathbf{v}_{\bar{\mathbf{x}}})
&=&v_{\hat{x}_i}^{-1}-v_i^{-1}\nonumber\\
&\approx&  {v_{\hat{x}}(v)}^{-1}-v^{-1} \nonumber\\
&=& \frac{1}{N_u} \mathrm{Tr}\{\left(\sigma _{{{n}}}^{- 2}w^2\mathbf{H}^H\mathbf{H}+v^{-1}\mathbf{I}_{N_u}\right)^{-1}\}^{-1} - v^{-1}\nonumber\\
&=&\phi(v)\nonumber\\
&=&\rho.
\end{eqnarray}

The approximations here is based on the symmetry of the system model and the law of large numbers. Similarly, we have $\psi_i(\rho_i)=\psi(\rho)$, $i\in \mathcal{N}_u$, and the following matching conditions

\emph{Proposition 4}: \emph{For any $i\in \mathcal{N}_u$, the matching conditions of the iterative symmetric MIMO-NOMA system can be rewritten as}
\begin{eqnarray}
\psi(\rho)&=&\phi^{-1}(\phi({1}))=1, \;\;\mathrm{for}\;\; 0\leq\rho<\phi({1});\label{e39}\\
\psi(\rho)&=&\phi^{-1}(\rho),\;\;\mathrm{for}\;\; \phi({1})\leq\rho<\phi({0});\label{e40}\\
\psi(\rho)&=&0\;\;\mathrm{for},\;\; \phi({0})\leq\rho<\infty.\label{e41}
\end{eqnarray}

Therefore, the analysis of the transfer functions for the MIMO-NOMA systems are degenerated into that of the single-user and single-antenna systems. Then, we can get the following theorem.

\textbf{\emph{Theorem 1}}: \emph{For a symmetric Large-scale MIMO-NOMA system whose users have: (i) the same rate $R_i=R$ for $i\in \mathcal{N}_u$; (ii) the same power $K_{\mathbf{x}}=w^2I$, the Iterative LMMSE detection achieves the capacity, i.e., $R_{i}=\frac{1}{N_{u}}\log \det\left(I_{N_r} + \frac{w^2}{\sigma_n^2}\mathbf{H}\mathbf{H}^H\right)$ for $i\in \mathcal{N}_u$ and $R_{sum}=\log \det\left(I_{N_r} + \frac{w^2}{\sigma_n^2}\mathbf{H}\mathbf{H}^H\right)$.}

\begin{proof}
Theorem 1 can be proved either by (\ref{e35}) at the decoder or by (\ref{e36}) at the estimator directly. Here, we show the first proof, and the other one is given in APPENDIX A.

\emph{i) User Rate Calculation:} From (\ref{e35}), the achievable rate of user $i$ is given by
\begin{eqnarray}\label{e42}
\!\!\!\!\!\!\!\!\!\!\!\!\!\!\!\!\!\!\!\!\!\!\!R_i&=&\int\limits_0^\infty  {{{\left( {{\rho _i} + {\psi _i}{{({\rho _i})}^{ - 1}}} \right)}^{ - 1}}d{\rho _i}}\nonumber\\
&\mathop  \approx \limits^{(a)} &\int\limits_0^\infty  {{{\left( {{\rho } + {\psi }{{({\rho })}^{ - 1}}} \right)}^{ - 1}}d{\rho }}\nonumber\\
&\mathop \leq \limits^{(b)}&\int\limits_{\phi(1)}^{\phi(0)}  {{\left[ {\rho } +  \left( {\phi }^{ - 1}{({\rho })}\right)^{ - 1} \right]}^{ - 1}d{\rho }} + \int\limits_{0}^{\phi(1)} {(1+\rho)^{-1}d\rho}  \nonumber\\
&\mathop  = \limits^{(c)}&\int\limits_{v=1}^{v=0}  {\left( v^{-1} +  \phi(v) \right)^{ - 1}d{\phi(v) }} + \log\left(1+\phi(v)\right)  \nonumber\\
&\mathop  = \limits^{(d)}&\int\limits_{v=1}^{v=0}  { v_{\hat{x}}(v) d {v_{\hat{x}}(v)}^{-1}} - \int\limits_{v=1}^{v=0}  {v_{\hat{x}}(v)d{v^{-1} }} - \log v_{\hat{x}}(v=1)\qquad\qquad\qquad\qquad\qquad\qquad\quad  \nonumber
\end{eqnarray}
\begin{eqnarray}
&\mathop  = \limits^{(e)}& - \int\limits_{v=1}^{v=0} {\frac{1}{N_u} \mathrm{Tr}\{\left(\sigma _{{{n}}}^{- 2}w^2\mathbf{H}^H\mathbf{H}+v^{-1}\mathbf{I}_{N_u}\right)^{-1}\} d{v^{-1} }} -\mathop {\lim }\limits_{v \to 0} \; \log {v_{\hat x}}(v) \nonumber\\
&\mathop  = \limits^{(f)}& \left[{ \frac{1}{N_u}\log \det\left( v^{-1}\mathbf{I}_{N_u} + \frac{w^2}{\sigma _n^{2}}\mathbf{H}^H\mathbf{H} \right)}\right]_{v=1}^{v=0} -\mathop {\lim }\limits_{v \to 0} \; \log \frac{1}{N_u} \mathrm{Tr}\{\left(\frac{w^2}{\sigma _{{{n}}}^{ 2}}\mathbf{H}^H\mathbf{H}+v^{-1}\mathbf{I}_{N_u}\right)^{-1} \nonumber\\
&=& { \frac{1}{N_u}\log \det\left( \mathbf{I}_{N_u} + {\frac{w^2}{\sigma_n^2}}\mathbf{H}^H\mathbf{H} \right) } \nonumber\\
&\mathop  = \limits^{(g)}& { \frac{1}{N_u}\log \det\left( \mathbf{I}_{N_r} + \frac{w^2}{\sigma_n^2}\mathbf{H}\mathbf{H}^H \right) } \nonumber\\
&=& R.
\end{eqnarray}
The approximation $(a)$ is based on $\rho_i=\rho$ and $\psi_i(\rho_i)=\psi(\rho)$, equations $(c)$ and $(d)$ are given by (\ref{e38}), equation $(e)$ comes from  (\ref{e37}), equation $(f)$ is based on the law $\int {\mathrm{Tr}\{ s\mathbf{I} + \mathbf{A}\} ds }= \log\det (s\mathbf{I} + \mathbf{A})$ and equation $(g)$ is derived by $\det(\mathbf{I}_K+\mathbf{A}_{K\times M} \mathbf{B}_{M\times K}) = \det(\mathbf{I}_M+\mathbf{B}_{M\times K}\mathbf{A}_{K\times M})$ for any matrixes $\mathbf{A}_{K\times M}$ and $\mathbf{B}_{M\times K}$. The inequality $(b)$ is derived by the matching condition (\ref{e39})$\sim$(\ref{e41}) and the equality holds if and only if there exists that code whose transfer function satisfies the matching condition. In the following, we show the existence of such code whose transfer function matches the transfer function of the LMMSE estimator.

\emph{ii) Code Existence:} From the ``\emph{Property of SCM Codes}", we can see that there exist such $n$-layer SCM codes exist if their transfer functions satisfy (i)$\sim$(iv) and $n$ is large enough. Therefore, in order to show the existence of such code, we only need to verify whether the matched transfer function meets the conditions (i)$\sim$(iv) . It is easy to see that conditions (i) and (iv) are always satisfied by (\ref{e39}) and (\ref{e41}) respectively. From (\ref{e38})$\sim$(\ref{e41}), we can see that $\psi(\rho)$ is continuous and differentiable in $[0,\infty)$ except at $\rho=\phi(0)$ and $\rho=\phi(1)$. Thus, Condition (iii) is satisfied.
To show the monotonicity of the transfer function, we rewrite (\ref{e38}) by the \emph{Matrix Inversion Lemma}.
\begin{eqnarray}\label{er}
\phi(v)&=&\frac{1}{{{N_u}}}\sum\limits_{i = 1}^{{N_u}} {\left[ {{{\left\{ {v - {v^2}\frac{{{w^2}}}{{\sigma _n^2}}\mathbf{h}_i^H{{\left( {{{\bf{I}}_{{N_r}}} + \frac{{{w^2}v}}{{\sigma _n^2}}{\bf{H}}{{\bf{H}}^H}} \right)}^{ - 1}}{\mathbf{h}_i}} \right\}}^{ - 1}} - {v^{ - 1}}} \right]} \nonumber \\
 &=& \frac{1}{{{N_u}}}\sum\limits_{i = 1}^{{N_u}} {{1 \mathord{\left/
 {\vphantom {1 {\left[ {{{\left[ {\frac{{{w^2}}}{{\sigma _n^2}}\mathbf{h}_i^H{{\left( {{v^{ - 1}}{{\bf{I}}_{{N_r}}} + \frac{{{w^2}}}{{\sigma _n^2}}{\bf{H}}{{\bf{H}}^H}} \right)}^{ - 1}}{\mathbf{h}_i}} \right]}^{ - 1}} - 1} \right]}}} \right.
 \kern-\nulldelimiterspace} {\left[ {{{\left( {\frac{{{w^2}}}{{\sigma _n^2}}\mathbf{h}_i^H{{\left( {{v^{ - 1}}{{\bf{I}}_{{N_r}}} + \frac{{{w^2}}}{{\sigma _n^2}}{\bf{H}}{{\bf{H}}^H}} \right)}^{ - 1}}{\mathbf{h}_i}} \right)}^{ - 1}} - 1} \right]}}} \nonumber\\
 &=& \frac{1}{{{N_u}}}\sum\limits_{i = 1}^{{N_u}} {{1 \mathord{\left/
 {\vphantom {1 {\left( {f_i^{ - 1}(v) - 1} \right)}}} \right.
 \kern-\nulldelimiterspace} {\left( {f_i^{ - 1}(v) - 1} \right)}}},
\end{eqnarray}
where $f_i(v)={\frac{{{w^2}}}{{\sigma _n^2}}\mathbf{h}_i^H{{\left( {{v^{ - 1}}{{\bf{I}}_{{N_r}}} + \frac{{{w^2}}}{{\sigma _n^2}}{\bf{H}}{{\bf{H}}^H}} \right)}^{ - 1}}{\mathbf{h}_i}}$. It is easy to verify that $f_i(v)$ is a decreasing function with respect to $v$, and $\phi(v)$ is thus a decreasing function of $v$. With the definition of $\psi(\rho)$ basded on (\ref{e39})$\sim$(\ref{e41}), we then see that $\psi(\rho)$ is a monotonically decreasing function in $[0,\infty)$. Therefore, the matched transfer function can be constructed by the SCM code, i.e., there exists such codes that satisfy the matching conditions.

\emph{iii) Sum Rate Calculation:} Thus, based on (i)$\sim$(iii), we get the achievable sum rate
\begin{eqnarray}\label{e44}
R_{sum}&=&\sum\limits_{i = 1}^{{N_u}} {{R_i}} \nonumber\\
 &\approx& {N_u}R \nonumber\\
&=& \log \det\left( \mathbf{I}_{N_r} + \frac{w^2}{\sigma _n^{ 2}}\mathbf{H}\mathbf{H}^H \right),
\end{eqnarray}
which is the exact sum capacity of the system. Therefore, we get the Theorem 1.
\end{proof}

Theorem 1 shows that for a symmetric large-scale MIMO-NOMA system, the iterative detection structure is optimal, i.e., the LMMSE estimator is an optimal estimator without losing any useful information during the estimation. Although this result is derived based on the large-scale MIMO-NOMA systems, the next achievable rate analysis for the asymmetric systems shows that it also works for the general asymmetric systems.

\subsection{Sum Capacity Achieving of Iterative LMMSE Detector for Asymmetric MIMO-NOMA Systems}

For general asymmetric MIMO-NOMA system, the achievable rate region calculation of the Iterative LMMSE detection is more complicated. As we mentioned, on the one hand, all the users' transfer functions interact with each other at the estimator, i.e., the any output of the estimator relies on every variance of the input messages from the decoders. On the other hand, the transfer curve of each decoder should lies below the transfer curve of the estimator. The estimator and decoders are associated with each other. It is hard to try all the possible available transfer functions of the decoders to get the optimal code design of each user. However, the area theorem tells us that the achievable rate of every user is maximized if and only if its transfer function matches with that of the estimator if there exists some codes with that transfer function. Therefore, we can arbitrarily choose an input variances of the estimator from the decoders, and then achieve the transfer functions of the estimator. As a result, we get users' achievable rate by matching the decoders' transfer functions with the estimator.
\begin{figure}[ht]
  \centering
  \includegraphics[width=12cm]{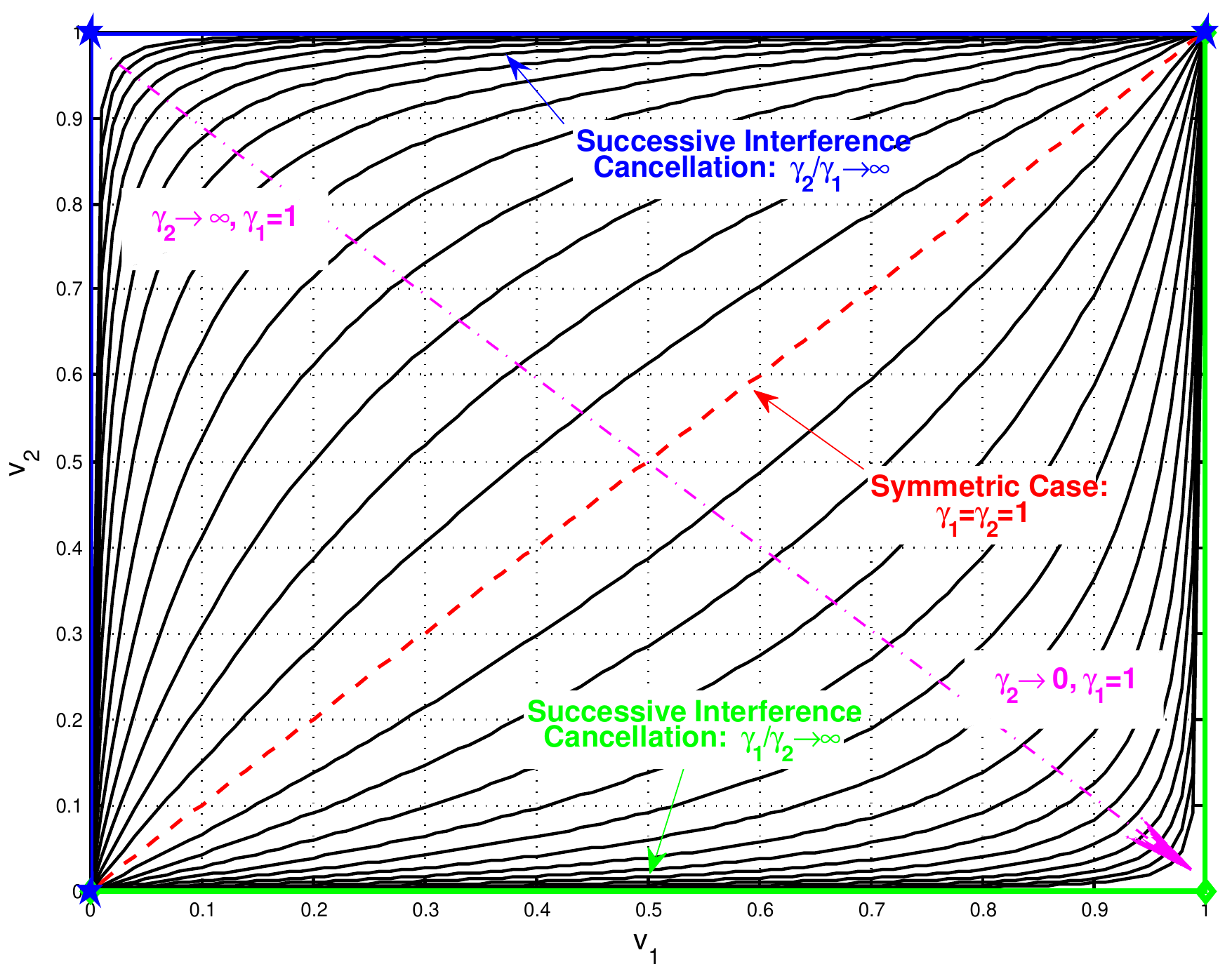}\\\vspace{-0.4cm}
  \caption{Variance tracks for the different $\bm{\gamma}$, where $\gamma_1=1$ is fixed. The variance of user $i$ is $v_i$, $i=1,2$. When $\gamma_2$ changes from $\infty$ to $0$, the track change from the blue curve (SIC case and the decoding order is user $1\to$ user $2$) to green curve (SIC case and the decoding order is user $2\to$ user $1$). When $\gamma_1=\gamma_2=1$, it is degenerated to the symmetric case (red line). }\label{f3}\vspace{-0.4cm}
\end{figure}
\begin{figure}[ht]
  \centering
  \includegraphics[width=15cm]{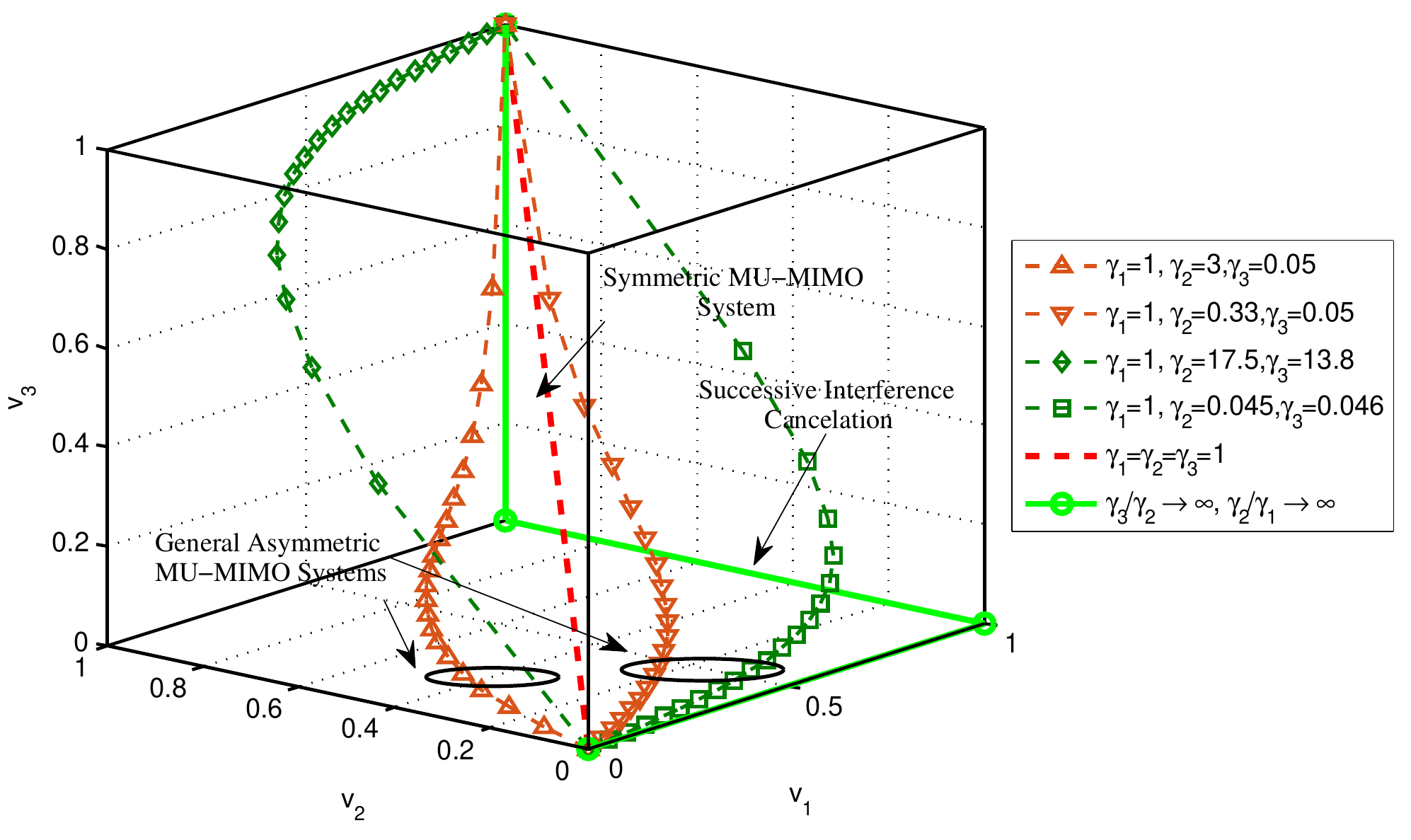}\\\vspace{-0.4cm}
  \caption{Variance tracks for the different $\bm{\gamma}$, where $\gamma_1=1$ is fixed. The variance of user $i$ is denoted as $v_i$, $i=1,2,3$. The variance track changes with $\gamma_2$ and $\gamma_3$. When $\gamma_3/\gamma_2\to\infty$ and $\gamma_2/\gamma_1\to\infty$ (green curve),  it is degenerated to a SIC case with the decoding order: user $3\to$user $2\to$ user $1$. When $\gamma_1=\gamma_2=\gamma_3=1$, it is degenerated to the symmetric case (red line). The other curves are the general asymmetric cases.}\label{f4}\vspace{-0.4cm}
\end{figure}

To simplify the analysis, we add some linear constraints for the input variances of the ESE.
\begin{equation}\label{e45}
\gamma_i (v_i^{-1}-1)=\gamma_j (v_j^{-1}-1), \;\; \mathrm{for \;\; any} \;\;i,j\in \mathcal{N}_u.
\end{equation}
Without loss of generality, we assume $\gamma_1=1$ and $\gamma_i>0$ , that is, $v_i^{-1}=1+\gamma_i^{-1}(v_1^{-1}-1)$ for any $i\in \mathcal{N}_u$. Actually, the different $\bm{\gamma} = [\gamma_1, \cdots, \gamma_{N_u}]$ values give the different variance track during the iteration. Fig. \ref{f3} and Fig. \ref{f4} presents the variance tracks of the different $\bm{\gamma}$ for the two users and three users cases respectively. As we can see, when (\ref{e45}) concluded the symmetric case (when $w_1=\cdots =w_{N_u}$) and all the SIC points (maximal extreme points of the capacity region). If $\gamma_{k_i}/\gamma_{k_{i-1}}\to \infty$, for any $i\in \mathcal{N}_u/\{1\}$, we can get the SIC points with the decoding order $[k1,k2,\cdots,k_{N_u}]$, which is a permutation of $[1,2,\cdots,N_u]$. The blue curve and green curves in Fig. \ref{f3} and Fig. \ref{f4} are corresponding to some of the maximal extreme points. We will also show that the different variance tracks are corresponding to the different achievable rates of the users, i.e., the user's achievable rate can be adjusted by the parameter $\bm{\gamma}$.

With (\ref{e45}), we have
\begin{equation}\label{e46}
\mathbf{V}_{\bar{\mathbf{x}}}^{-1}= \mathbf{I}_{N_u} + \gamma_i(v_i^{-1}-1)\bm{\Lambda}_{\bm{\gamma}}^{-1} = \mathbf{V}_{\bar{\mathbf{x}}}^{-1}(v_i)
\end{equation}
and
\begin{eqnarray}\label{evall}
\mathbf{V}_{\hat{\mathbf{x}}}&=&(\sigma _{{{n}}}^{- 2}\mathbf{H}'^H\mathbf{H}'+\mathbf{V} _{{{ \bar{\mathbf{x}}}}}^{-1})^{-1}\nonumber\\
&=& (\sigma _{{{n}}}^{- 2}\mathbf{H}'^H\mathbf{H}'+ \mathbf{V}_{\bar{\mathbf{x}}}^{-1}(v_i))^{-1}\nonumber\\
&=& \mathbf{V}_{\hat{\mathbf{x}}}(v_i)
\end{eqnarray}
where $i\in\mathcal{N}_u$, $\bm{\Lambda}_{\bm{\gamma}}=\mathrm{diag}(\bm{\gamma})$ is a diagonal matrix whose diagonal entries are $\bm{\gamma}$. Thus, we have
\begin{equation}\label{ephi}
\phi_i(\mathbf{v}_{\bar{\mathbf{x}}})= v_{\hat{x}_i}(v_i)^{-1}-v_i^{-1}=\phi_i(v_i)=\rho_i,
\end{equation}
For example, if we take $i=1$, we have
\begin{equation}\label{ev_1}
\mathbf{V}_{\bar{\mathbf{x}}}^{-1}= \mathbf{V}_{\bar{\mathbf{x}}}^{-1}(v_1),\;\;
\mathbf{V}_{\hat{\mathbf{x}}} =  \mathbf{V}_{\hat{\mathbf{x}}}(v_1),\;\;
\phi_i(\mathbf{v}_{\bar{\mathbf{x}}})= \phi_i(v_1).
\end{equation}

\emph{ Proposition 5: Based on (\ref{ephi}), \emph{for $i\in \mathcal{N}_u$, the matching condition (\ref{e29}) can be rewritten to}
\begin{eqnarray}
\psi_i(\rho_i)&=&\phi_i^{-1}(\phi_i({1}))=1, \;\;\mathrm{for}\;\; 0\leq\rho_i<\phi_i({1});\label{em1}\\
\psi_i(\rho_i)&=&\phi_i^{-1}(\rho_i),\;\;\mathrm{for}\;\; \phi_i({1})\leq\rho_i<\phi_i({0});\label{em2}\\
\psi_i(\rho_i)&=&0,\;\;\mathrm{for}\;\; \phi_i({0})\leq\rho_i<\infty.\label{em3}
\end{eqnarray}}

Then, we can give the users' achievable rates of the Iterative LMMSE detection for the asymmetric MIMO-NOMA systems by the following lemmas.

\emph{\textbf{Lemma 1}: For the asymmetric MIMO-NOMA systems with Iterative LMMSE detection, the achievable rates of the users are
\begin{equation}\label{lemma1}
R_i = \int\limits_{v_1=1}^{v_1=0} \left[v_1- \gamma_i^{-1} \left[\mathbf{V}_{\hat{\mathbf{x}}}(v_1)\right]_{i,i} \right] dv_1^{-1}
- \log(\gamma_i),
\end{equation}
where $i\in \mathcal{N}_u$, $\mathbf{V}_{\hat{\mathbf{x}}}(v_1)=\left( \sigma^{-2}_n\mathbf{H}'^H\mathbf{H}'+ \mathbf{I}_{N_u} + (v_1^{-1}-1)\bm{\Lambda}_{\bm{\gamma}}^{-1}\right)^{-1}$, and $[\cdot]_{i,i}$ denotes the $i$-th column and $i$-th row entry of the corresponding matrix.}

\begin{proof}
Similarly, in this case, the achievable rate of user $i$ can be given either by (\ref{e35}) at the decoder or by (\ref{e36}) at the estimator directly. Here, we give the first proof, and the other one is given in APPENDIX B. From (\ref{e35}), the achievable rate of user $i$ is given as
\begin{eqnarray}\label{e47}
R_i&=&\int\limits_0^\infty  {{{\left( {{\rho _i} + {\psi _i}{{({\rho _i})}^{ - 1}}} \right)}^{ - 1}}d{\rho _i}}\nonumber\\
&\mathop \leq \limits^{(a)}&\int\limits_{\phi_i(1)}^{\phi_i(0)}  {{\left[ {\rho_i } +  \left( {\phi_i }^{ - 1}{({\rho_i })}\right)^{ - 1} \right]}^{ - 1}d{\rho_i }} + \int\limits_{0}^{\phi_i(1)} {(1+\rho_i)^{-1}d\rho_i}  \nonumber\\
&\mathop = \limits^{(b)}&\int\limits_{v_i=1}^{v_i=0}  {\left( v_i^{-1} +  \phi_i(v_i) \right)^{ - 1}d{\phi_i(v_i) }} + \log\left(1+\phi_i(v_i)\right)  \nonumber\\
&\mathop = \limits^{(c)}&\int\limits_{v_i=1}^{v_i=0}  { v_{\hat{x}_i}(v_i) d {v_{\hat{x}_i}(v_i)}^{-1}} - \int\limits_{v_i=1}^{v_i=0}  {v_{\hat{x}_i}(v_i)d{v_i^{-1} }} - \log v_{\hat{x}_i}(v_i=1)  \nonumber\\
&\mathop = \limits^{(d)}& - \int\limits_{v_1=1}^{v_1=0} { \gamma_i^{-1} \left[\mathbf{V}_{\hat{\mathbf{x}}}(v_1)\right]_{i,i} dv_1^{-1}}
-\mathop {\lim }\limits_{v_1 \to 0} \; \log\left[\mathbf{V}_{\hat{\mathbf{x}}}(v_1)\right]_{i,i}\;\nonumber\\
&\mathop = \limits^{(e)}& \int\limits_{v_1=1}^{v_1=0} \left[v_1- \gamma_i^{-1} \left[\mathbf{V}_{\hat{\mathbf{x}}}(v_1)\right]_{i,i} \right] dv_1^{-1} - \log(\gamma_i).
\end{eqnarray}
The inequality $(a)$ is derived by (\ref{em1})$\sim$(\ref{em3}) and the equality holds if and only if there exists that code whose transfer function satisfies the matching condition. The equations $(b)\sim(d)$ are given by $\rho_i=\phi_i(v_i)$, (\ref{ephi}) and (\ref{ev_1}), equation $(e)$ comes from  (\ref{e46}) and (\ref{evall}).

In APPENDIX C, similar to the Theorem 1, we show the existence of such codes whose transfer functions match the transfer functions of the LMMSE estimator. In APPENDIX D, the existence of the infinite integral of (\ref{lemma1}) is proved.
\end{proof}

\emph{\textbf{Lemma 2}: The achievable rate $R_i$ of user $i$ increases monotonously with $\gamma_i$ and decreases monotonously $\gamma_j$, where $i,j \in \mathcal{N}_u$ and $j\neq i$.}
\begin{IEEEproof}
It is easy to find that $\mathrm{mmse}^{ese}_{tot,i}$ (or $\mathrm{mmse}^{dec}_{tot,i}$) increases monotonously with $\gamma_i$ and decreases monotonously $\gamma_j$ for $i,j \in \mathcal{N}_u$ and $j\neq i$. Thus, based on the \emph{Proposition 2}, we have that $R_i$ increases monotonously with $\gamma_i$ and decreases monotonously $\gamma_j$ for $j\neq i$.
\end{IEEEproof}

It should be noted that although the \emph{Lemma 1} gives the achievable rate of the users and it is an integral function of channel matrix, noise variance and $\Lambda_{\bm{\gamma}}$, we cannot see the specific relationship between the achievable rates and $\Lambda_{\bm{\gamma}}$ because of the complicated integral structure of (\ref{lemma1}). Therefore, it is very hard to show the analytical achievable rate region of the Iterative LMMSE detection. However, the sum capacity achieving of the Iterative LMMSE detection can be shown by the following theorem.

\emph{\textbf{Theorem 2:} The Iterative LMMSE detection achieves the sum capacity of the MIMO-NOMA systems, i.e., $R_{sum}=\log \det\left(I_{N_u} + {\sigma_n^{-2}}\mathbf{H}'\mathbf{H}'^H\right)$.}

\begin{IEEEproof}
With (\ref{lemma1}), the achievable sum rate is
\begin{eqnarray}\label{e49}
R_{sum}&=&\sum\limits_{i = 1}^{{N_u}} {{R_i}} \nonumber\\
 &\mathop\leq \limits^{(a)}&- \int\limits_{v_1=1}^{v_1=0} \sum\limits_{i = 1}^{{N_u}} {{ \left(\gamma_i^{-1} \left[\mathbf{V}_{\hat{\mathbf{x}}}(v_1)\right]_{i,i} \right) dv_1^{-1}}  }
-\mathop {\lim }\limits_{v_1 \to 0} \; \log(v_1^{N_u}\mathop \Pi \limits_{i = 1}^{{N_u}} \gamma_i) \nonumber\\
 &=&- \int\limits_{v_1=1}^{v_1=0}  {{ \mathrm{Tr}\{\bm{\Lambda}_{\bm{\gamma}}^{-1} \mathbf{V}_{\hat{\mathbf{x}}}(v_1) \} dv_1^{-1}}  }
-\mathop {\lim }\limits_{v_1 \to 0} \; \log(v_1^{N_u}\mathop \Pi \limits_{i = 1}^{{N_u}} \gamma_i) \nonumber\\
 &\mathop = \limits^{(b)}&-\mathop {\lim }\limits_{v_1 \to 0} \; \log(v_1^{N_u}\mathop \Pi \limits_{i = 1}^{{N_u}} \gamma_i)- \left[ \log\det\left( (v_1^{-1}-1)\mathbf{I}_{N_u} + \left( \mathbf{I}_{N_u} + \sigma_n^{-2}\mathbf{H}'^H\mathbf{H}' \right)\bm{\Lambda}_{\bm{\gamma}} \right)\right]_{v_1=1}^{v_1=0}
 \nonumber\\
  &=&-\mathop {\lim }\limits_{v_1 \to 0} \; \log(v_1^{N_u}\mathop \Pi \limits_{i = 1}^{{N_u}} \gamma_i)- \mathop {\lim }\limits_{v_1 \to 0} {\log \det(v_1^{-1}\mathbf{I}_{N_u}) } + \log \det\left( \left(\mathbf{I}_{N_u} + {\sigma _n^{- 2}}\mathbf{H}'^H\mathbf{H}'\right)\bm{\Lambda}_{\bm{\gamma}} \right)
 \nonumber\\
&=& \log \det\left( \mathbf{I}_{N_u} + {\sigma _n^{- 2}}\mathbf{H}'^H\mathbf{H}' \right),
\end{eqnarray}
which is the exact system sum capacity. The inequality $(a)$ is derived by (\ref{e48}) or (\ref{e49}), and equation $(b)$ is based on (\ref{e48}) (\ref{evall}) and the law $\int {\mathrm{Tr}\{ \left(s\mathbf{I} + \mathbf{A}\right)^{-1}\} ds }= \log\det (s\mathbf{I} + \mathbf{A})$. It means that the iterative detector is sum capacity-achieving for the different kinds of user-rate combinations.
\end{IEEEproof}

Theorem 2 shows that for a general asymmetric MIMO-NOMA system, from the sum rate perspective, the iterative detection structure is optimal, i.e., the LMMSE estimator is an optimal estimator without losing any useful information during the estimation.

\textbf{\emph{Remark 1:}} When $\gamma_i=1$ for $i\in \mathcal{N}_u$, and for a symmetric system with: \emph{(i)} the same rate $R_i=R$ for $i\in \mathcal{N}_u$; \emph{(ii)} the same power $K_{\mathbf{x}}=w^2I$, \emph{Theorem 2} can be degenerated to the \emph{Theorem 1}. Thus, the symmetric system is a special case of the asymmetric systems.

\vspace{-0.1cm}\begin{algorithm}
\caption{Numeric Iterative Search Algorithm for $\Lambda_{\bm{\gamma}}$}
\begin{algorithmic}[1]
\State {\small{\textbf{Input:} {{$\mathbf{H}$,  $K_{\mathbf{x}}$, $\sigma^2_n$, $\epsilon>0$, $N_{max}$}}, $\mathbf{R}=[R_1,\cdots,R_{N_u}]$ and calculate {{$\mathbf{H}'$}}.
\State \textbf{If} $\mathbf{R}\in \mathbf{\mathcal{R}}_\mathcal{S}$ ($\mathbf{\mathcal{R}}_\mathcal{S}$ is the capacity region given by (\ref{e17}))
\State \quad\;\; \textbf{Initialize:} Random choose $\bm{\gamma}=[\gamma_1,\cdots,\gamma_{N_u}]$, $\gamma_i>0$ for $\forall i\in \mathcal{N}_u$,
\Statex \quad\quad\quad\quad\quad\quad\; Calculate $\mathbf{R}^{(0)}(\bm{\gamma})=[R_1^{(0)},\cdots,R_{N_u}^{(0)}]$ by (\ref{lemma1}) and $t=1$.
\State \quad\;\; \textbf{While} \;{\small{$\left( \:||\mathbf{R}^{(0)}-\mathbf{R}||_1>\epsilon \;{\textbf{or}}\; t<N_{max} \;\right)$}}
\State \quad\quad\;\; \textbf{For} $i=1:N_u$
\State \quad\quad\quad\;\; fixed $\bm{\gamma}_{\sim i}=[\gamma_1,\cdots,\gamma_{i-1},\gamma_{i-1}, \cdots,\gamma_{N_u}]$, \Statex \quad\quad\quad\;\; find $\gamma_i^*$ for ${R}_i^{(1)}(\gamma_i=\gamma_i^*)={R}_i$, and
\State \quad\quad\quad\;\; calculate $\mathbf{R}^{(1)}(\bm{\gamma}_{\sim i},\gamma_i^*)=[R_1^{(1)},\cdots,R_{N_u}^{(1)}]$.
\State \quad\quad\quad\;\; \textbf{While} $ ||\mathbf{R}^{(1)}-\mathbf{R}||_1>||\mathbf{R}^{(0)}-\mathbf{R}||_1$ \State \quad\quad\quad\quad\quad\;\;$\gamma_i^*=(\gamma_i+\gamma_i^*)/2$ and go to step 7.
\State \quad\quad\quad\;\; \textbf{End While}
\State \quad\quad\quad\;\; $\gamma_i=\gamma_i^*$ and $\mathbf{R}^{(0)}=\mathbf{R}^{(1)}$.
\State \quad\quad\;\; \textbf{End For}
\State \quad\quad\;\; $t=t+1$.
\State \quad\;\; \textbf{End While}
\State \quad\;\; \textbf{If} $t<N_{max}$
\State \quad\quad\;\; \textbf{Output:}  $\bm{\gamma}$.
\State \quad\;\; \textbf{Else}
\State \quad\quad\;\; go to step 21.
\State \quad\;\; \textbf{End If}
\State \textbf{Else} $\mathbf{R}\notin \mathbf{\mathcal{R}}_\mathcal{S}$
\State \quad\;\; \textbf{Output:} The given rate $\mathbf{R}$ is outside the capacity region or not achievable.
\State \textbf{End If} }}
\end{algorithmic}
\end{algorithm}

\subsection{Practical Iterative LMMSE Detection Design for Asymmetric MIMO-NOMASystems}
Actually, the proof of the codes existence for matching conditions also gives the process of optimal codes design for the MIMO-NOMA system. It should be noted that the codes design dependents on the realization of the channel matrix. Therefore, the users should know the information of the channel matrix. In addition, the codes design also depends on $\Lambda_{\bm{\gamma}}$. As we cannot get a closed-form solution of the user rates with respect to $\Lambda_{\bm{\gamma}}$, it is hard to find the proper $\Lambda_{\bm{\gamma}}$ for the given user rates. In this subsection, we propose an algorithm to search a numeric solution of $\Lambda_{\bm{\gamma}}$ to satisfy the rate requirement of each user.

Algorithm 1 gives a numeric iterative search of $\Lambda_{\bm{\gamma}}$ for the given rate $\mathbf{R}$, where $N_{max}$ is the maximum iterative number, $\epsilon$ indicates the allowed precision and $||\cdot||_1$ denotes the 1-norm. It should be noted that $\gamma^*_i$ in step 6 definitely exists and can be easy searched by dichotomy or quadratic interpolation method as $R_i$ increases monotonously with $\gamma_i$ (Lemma 2). In addition, steps $8\sim 10$ ensure that the new $\gamma_i^*$ always better than the previous one and the search program will not stop until the requirement $\Lambda_{\bm{\gamma}}$ is got. Experimentally, we find that the points in the system capacity region are always achievable by this numeric algorithm.

\section{Some Special Cases of Asymmetric MIMO-NOMA Systems}
In Section IV, we proved that the Iterative LMMSE detection achieves the sum capacity of the asymmetric systems, but whether it achieves the whole capacity region of the asymmetric MIMO-NOMA systems is still unkown. In this section, we analyse some special cases of the asymmetric MIMO-NOMA systems. We will show that: \emph{(i)} for the 2-user MIMO-NOMA system, it is proved that the Iterative LMMSE detection achieves the whole capacity region of the system, \emph{(ii)} all the maximal extreme points in the capacity region of the MIMO-NOMA system can be achieved by the Iterative LMMSE detection, and \emph{(iii)} for the 3-user MIMO-NOMA system, the simulation results show that the Iterative LMMSE detection can also achieve the whole capacity region of the MIMO-NOMA system.

\subsection{ Maximal Extreme Point Achieving of Iterative LMMSE Detection}
As it mentioned in the \emph{Capacity Region Domination Lemma}, the whole capacity region is dominated by a convex combination of the maximal extreme points, which has been proved that can be achieved by the SIC method. Here, we will show that all these maximal extreme points can be achieved by the Iterative LMMSE detection when the parameter $\Lambda_{\bm{\gamma}}$ are properly chosen.

\textbf{\emph{Corollary 1}}: \emph{All the maximal extreme points of the system capacity region can be achieved by the Iterative LMMSE detection.}

\begin{IEEEproof}
For any maximal extreme point expressed in (\ref{dominate}) with order vector [$k_1,\cdots,k_{N_u}]$, we let $\gamma_{k_i}/\gamma_{k_{i-1}}\to \infty$, for any $i\in \mathcal{N}_u/\{1\}$. Therefore, similar to the green curves showed in Fig. \ref{f3} and Fig. \ref{f4}, the user $k_{N_u}$ is recovered after all the variances of other users already being zeros as $\gamma_{k_{N_u}}/\gamma_{k_{i-1}}\to \infty$, for any $i\in \mathcal{N}_u/\{1\}$.

Thus, from (\ref{lemma1}), the rate of user $k_{N_u}$ is
\begin{equation}\label{coro1}
R_{k_{N_u}}= \log \left( {1 + \frac{1}{{\sigma _n^2}}{\mathbf{h}'}_{{k_{N_u}}}^H{{\mathbf{h}'}_{{k_{N_u}}}}} \right),
\end{equation}
which is the same as that in (\ref{dominate}).
Similarly, when we recovering the $k_{N_u-1}$, all the users have been recovered except user $k_{N_u}$ and user $k_{N_u}-1$. Base on this and \emph{Theorem 2}, we have
\begin{equation}\label{coro2}
R_{k_{N_u-1}} + R_{k_{N_u}} = \log\det\left(\mathbf{I}_{|\mathcal{S}_{N_u-2}^c|} + \frac{1}{\sigma_n^2} \mathbf{H}_{\mathcal{S}^c_{N_u-2}}'^H \mathbf{H}_{\mathcal{S}^c_{N_u-2}}'\right).
\end{equation}
Thus, based on (\ref{coro1}) and (\ref{coro2}), the rate of user $k_{N_u-1}$ is
\begin{equation}
R_{k_{N_u-1}}= \log\det\left(\mathbf{I}_{|\mathcal{S}_{N_u-2}^c|} + \frac{1}{\sigma_n^2} \mathbf{H}_{\mathcal{S}^c_{N_u-2}}'^H \mathbf{H}_{\mathcal{S}^c_{N_u-2}}'\right)-\log \left( {1 + \frac{1}{{\sigma _n^2}}{\mathbf{h}'}_{{k_{N_u}}}^H{{\mathbf{h}'}_{{k_{N_u}}}}} \right),
\end{equation}
which is the same as that in (\ref{dominate}). Continue this process and we can show all the other users' rates are the same as that of in (\ref{dominate}). Therefore, we have \emph{Corollary 1}.
\end{IEEEproof}

These corollary shows that as the parameter $\Lambda_{\bm{\gamma}}$ be properly chosen, the Iterative LMMSE detection can be degenerated to the SIC methods, i.e., the SIC methods are some special cases of the proposed Iterative LMMSE detection.

\subsection{ Capacity Region Achieving for Two-user MIMO-NOMA Systems}

As it is mentioned, it is very hard to calculate the specific achievable user rates from (\ref{lemma1}) for the general asymmetric MIMO-NOMA systems. In this subsection, we show that the Iterative LMMSE detection can achieve the whole capacity region of two-user MIMO-NOMA systems.

\emph{\textbf{Theorem 3}}: \emph{The Iterative LMMSE detection achieves the whole capacity region of two-user MIMO-NOMA systems as follows.}
\begin{equation}
\left\{ \begin{array}{l}
R_1\leq\log ( {1 + \frac{1}{{\sigma _n^2}}{\mathbf{h}'}_1^H{{\mathbf{h}'}_1}} ),\\
R_2\leq\log ( {1 + \frac{1}{{\sigma _n^2}}{\mathbf{h}'}_2^H{{\mathbf{h}'}_2}} ),\\
R_1+R_2\leq\log \det\left( \mathbf{I}_{2} + {\sigma _n^{- 2}}\mathbf{H}'^H \mathbf{H}'\right).
\end{array} \right.
\end{equation}

\begin{figure}[ht]
  \centering
  \includegraphics[width=12cm]{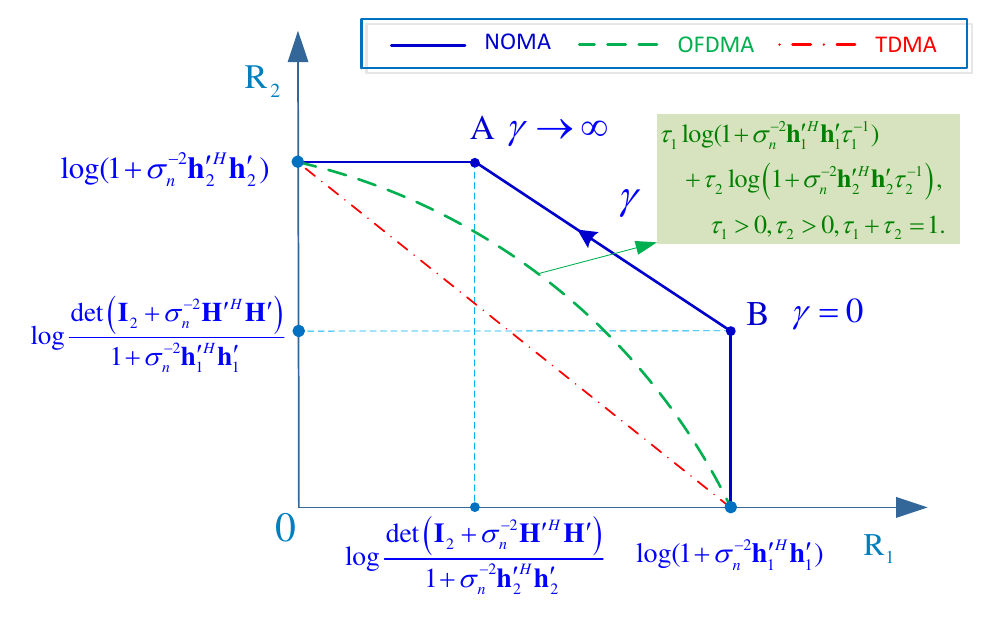}\\\vspace{-0.4cm}
  \caption{Capacity region achieving of Iterative LMMSE detector for two-user MIMO-NOMA system. When the parameter $\gamma$ changes from $0$ to $\infty$, point $(R_1,R_2)$ moves from maximal extreme point B to maximal extreme point A along segment AB. }\label{f5}\vspace{-0.1cm}
\end{figure}

\begin{IEEEproof}
The pentagon in Fig. \ref{f5} indicates the capacity region of two-user MIMO-NOMA system, which is dominated by segment AB and point A and point B are two maximal extreme points. Without loss of generality, we let $\gamma_1=1$ and $\gamma_2=\gamma\in[0,\infty)$. From \emph{Theorem 2}, we get
\begin{equation}\label{twouser1}
R_{sum}=R_1+R_2=\log \det\left( \mathbf{I}_{2} + {\sigma _n^{- 2}}\mathbf{H}'^H \mathbf{H}'\right),
\end{equation}
which is the exact sum capacity of the system.

In addition, as we discussed in \emph{Corollary 1}, when $\gamma$ changes from $0$ to $\infty$,  $R_1$ reduces from $\log \left( {1 + \frac{1}{{\sigma _n^2}}{\mathbf{h}'}_1^H{{\mathbf{h}'}_1}} \right)$ to $\log \det\left( \mathbf{I}_2 + {\sigma _n^{- 2}}\mathbf{H}'^H\mathbf{H}' \right)-\log \left( {1 + \frac{1}{{\sigma _n^2}}{\mathbf{h}'}_1^H{{\mathbf{h}'}_1}} \right)$, and $R_2$ increases from $\log \det\left( \mathbf{I}_2 + {\sigma _n^{- 2}}\mathbf{H}'^H\mathbf{H}' \right)-\log \left( {1 + \frac{1}{{\sigma _n^2}}{\mathbf{h}'}_2^H{{\mathbf{h}'}_2}} \right)$ to $\log \left( {1 + \frac{1}{{\sigma _n^2}}{\mathbf{h}'}_2^H{{\mathbf{h}'}_2}} \right)$. As the $R_1$ and $R_2$ are both continuous functions of $\gamma$, from (\ref{twouser1}), we can see that when the parameter $\gamma$ changes from $0$ to $\infty$, the point $(R_1,R_2)$ moves from maximal extreme point B to maximal extreme point A along the segment AB. It means that the Iterative LMMSE detection can achieve any point on the segment AB. Therefore, the Iterative LMMSE detection achieves the whole capacity region, because it is dominated by the segment AB.
\end{IEEEproof}

Actually, for the simple two-user case, we can give the specific expressions of $R_1$ and $R_2$. The following corollary is derived directly based on \emph{Lemma 1}.

\emph{\textbf{Corollary 2}: For two-user MIMO-NOMA system with Iterative LMMSE detection, the user rates are given by}\vspace{-0.3cm}
\begin{equation}\label{coro2_1}
\left\{ \begin{array}{l}
R_1=\frac{1}{2}\log(\gamma\det(A)) +\frac{a_{22}\gamma-a_{11}}{2\eta} \log\frac{a_{22}\gamma+a_{11}-\eta}{a_{22}\gamma+a_{11}+\eta},\\
R_2= \frac{1}{2}\log(\gamma^{-1}\det(A)) -\frac{a_{22}\gamma-a_{11}}{2\eta} \log\frac{a_{22}\gamma + a_{11}+\eta}{a_{22}\gamma+a_{11}+\eta},
\end{array} \right.
\end{equation}
where $\mathbf{A}=\sigma^{-2}_n\mathbf{H}'^H\mathbf{H}'+ \mathbf{I}_{2}=\left[  {\begin{array}{*{20}{c}}
  \vspace{-0.2cm} {{a_{11}}}&{{a_{12}}} \\
  {{a_{21}}}&{{a_{22}}}
\end{array}} \right]$ and $\eta=\sqrt {a_{22}^2{\gamma ^2} + 2(2{a_{21}}{a_{12}} - {a_{22}}{a_{11}})\gamma  + a_{11}^2} $. It is easy to find that $\eta$ is a real number as $\mathbf{A}$ is positive definite and $\gamma\geq0$.

\begin{figure}[ht]
  \centering
  \includegraphics[width=10cm]{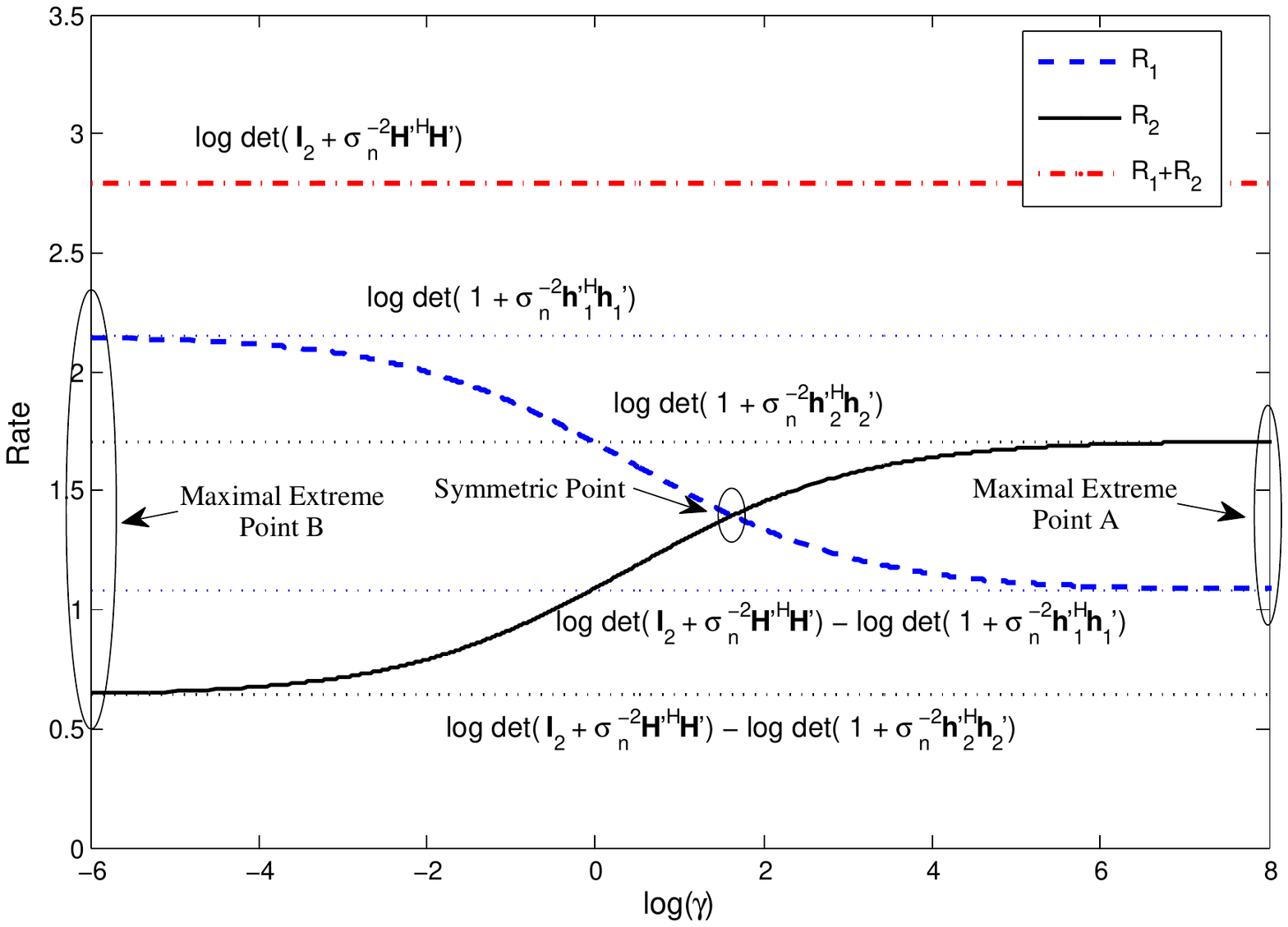}\\\vspace{-0.4cm}
  \caption{Relationship between the user rates and parameter $\gamma$ of the Iterative LMMSE detection for two-user MIMO-NOMA system. $N_r=2$, $N_u=2$, $\sigma_N^2=0.5$ and $\mathbf{H}=[
  1.32\;-1.31; \; -1.43 \;0.74]$. }\label{f6}\vspace{-0.4cm}
\end{figure}
\begin{figure}
  \centering
  \includegraphics[width=10cm]{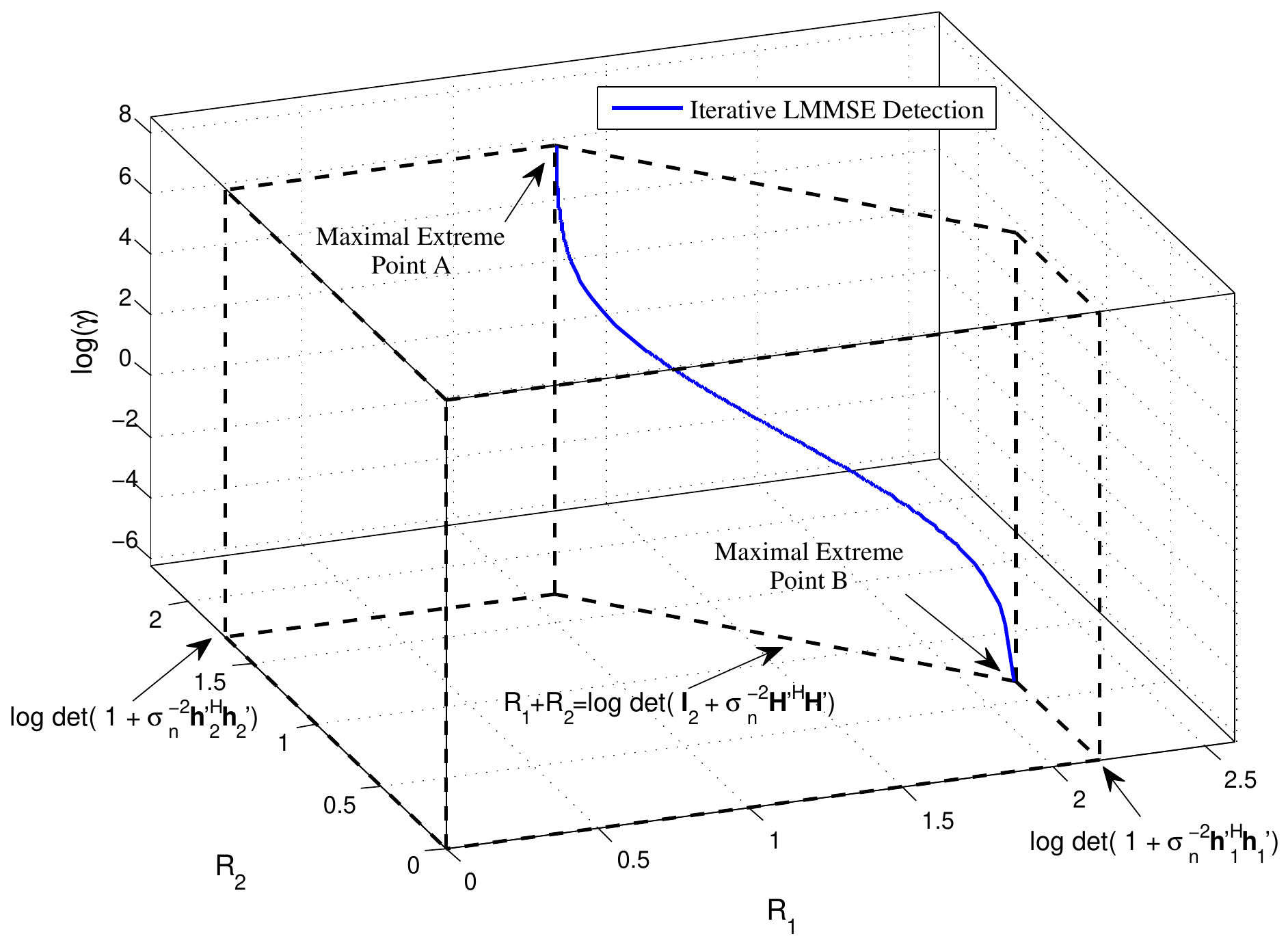}\\\vspace{-0.5cm}
  \caption{Relationship between the user rates and parameter $\gamma$ of the Iterative LMMSE detection for two-user MIMO-NOMA system. $N_r=2$, $N_u=2$, $\sigma_N^2=0.5$ and $\mathbf{H}=[
  1.32\; -1.31; \; -1.43\;0.74]$. }\label{f7}\vspace{-0.0cm}
\end{figure}

\textbf{\emph{Remark 2:}} It should be noted from (\ref{coro2_1}) that $R_1$ and $R_2$ are not linear functions of $\gamma$. It is easy to check that $R_1+R_2=\log \det\left( \mathbf{I}_{2} + {\sigma _n^{- 2}}\mathbf{H}'^H \mathbf{H}'\right)$, and when $\gamma\to0$ (or $\gamma\to\infty$), the limit of $(R_1,R_2)$ in (\ref{coro2_1}) converges to the maximal point B (or A) in Fig. \ref{f5}. When the parameter $\gamma$ changes from $0$ to $\infty$, the point $(R_1,R_2)$ can achieve any point on the segment AB in Fig. \ref{f5}. It also shows another proof of \emph{Theorem 3}. In addition, the achievable rates of TDMA and OFDMA are strictly smaller than that of the Iterative LMMSE NOMA systems. Fig. \ref{f6} and Fig. \ref{f7} present the rate curves of $R_1$ and $R_2$ respect to the parameter $\gamma$. It verifies that $R_2$ increases monotonously with the $\gamma$ and $R_1+R_2$ always equals to the system sum capacity.

\subsection{Capacity Region Achieving for Three-user MIMO-NOMA Systems}
\begin{figure}
      \centering
      \includegraphics[width=16cm]{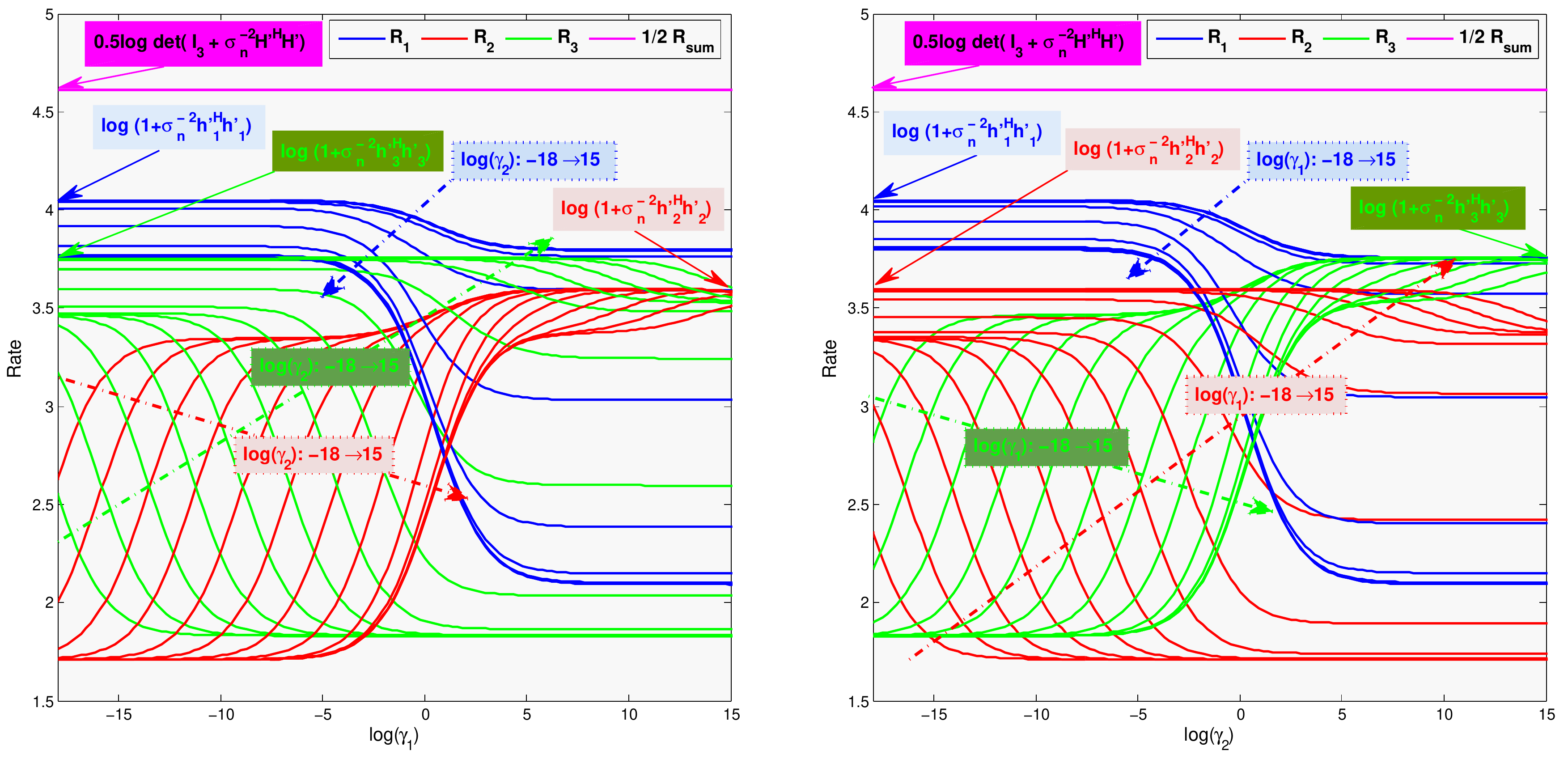}\\\vspace{-0.0cm}
      \caption{ Relationship between the user rates and parameters $(\gamma_1,\gamma_2)$ of the Iterative LMMSE detection for three-user MIMO system. $N_r=2$, $N_u=3$, $\sigma_N^2=0.5$ and $\mathbf{H}=[0.678 \;  0.603 \;  0.655;\; 0.557 \;  0.392  \;  0.171]$.}\label{f8}\vspace{-0.0cm}
\end{figure}
\begin{figure}
      \centering
      \includegraphics[width=16cm]{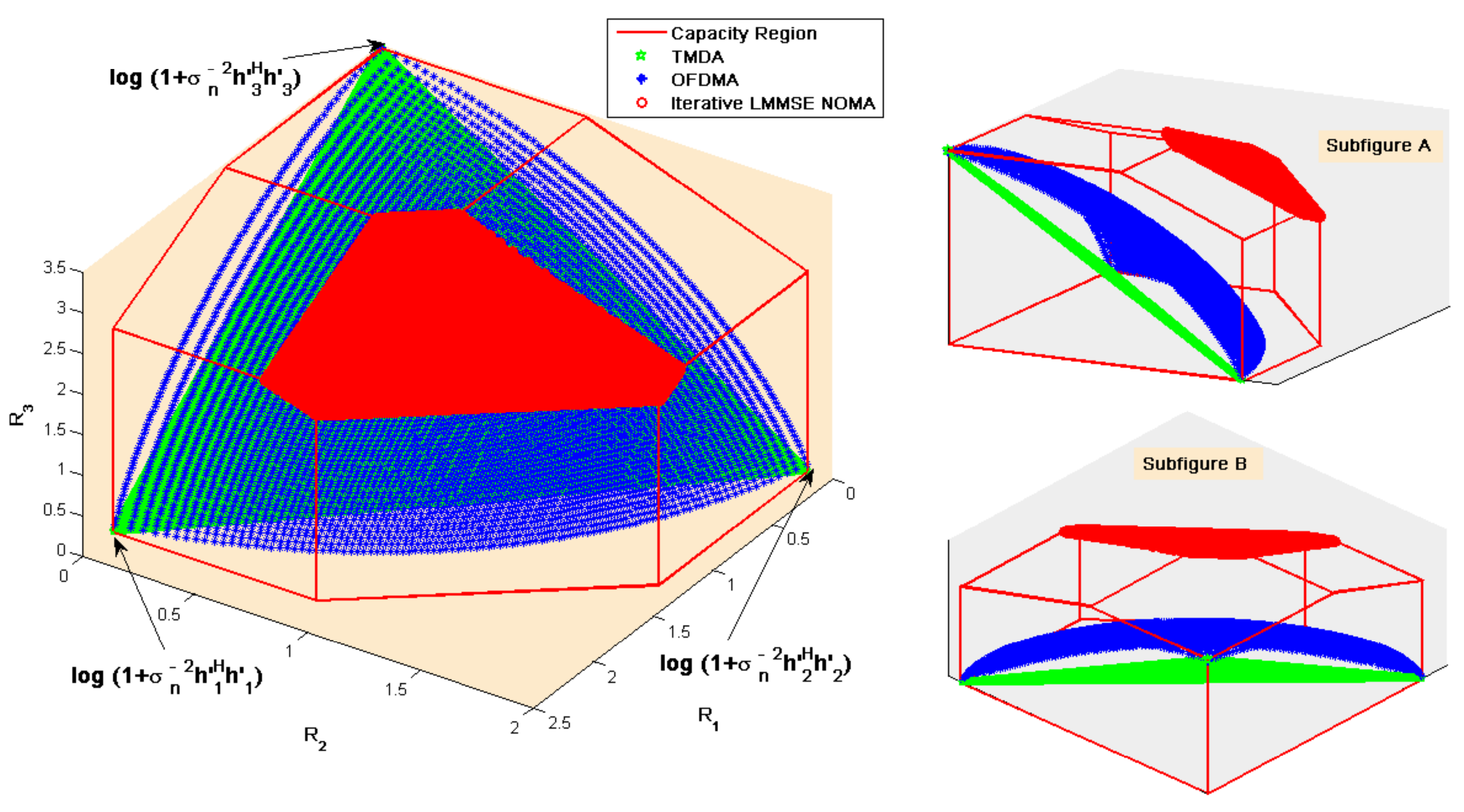}\\\vspace{-0.0cm}
      \caption{ Achievable user rates for all $(\gamma_1,\gamma_2)$ of the Iterative LMMSE detection for three-user MIMO-NOMA system. $N_r=3$, $N_u=3$, $\sigma_N^2=0.5$ and $\mathbf{H}=[1.95\;  1.28 \;  -2.53;\;
                 -0.31 \;  -0.16  \;  2.22;\; 0.55\; 1.08\; -1.98]$. Subfigure A and Subfigure B are the same figure with different ratated viewports.}\label{f9}\vspace{0cm}
\end{figure}

For the three-user MIMO-NOMA systems, it is hard to get a closed-form solution of the user rates. Therefore, it is difficult to show that the Iterative LMMSE detection can achieve the whole capacity region. However, the user rates can be solved by numerical calculation for (\ref{lemma1}). Fig. \ref{f8} shows the relationships between the user rates and $\gamma_1$ and $\gamma_2$, where $N_r=2$, $N_u=3$, $\sigma_N^2=0.5$, and $\mathbf{H}=[0.678 \;  0.603 \;  0.655;\; 0.557 \;  0.392  \;  0.171]$. It should be noted that although the user rates change with $\gamma_1$ and $\gamma_2$, the sum rate $R_{sum}=\sum\limits_{{\text{i}} = 1}^3 {{R_i}} $ is constant and equals to the system sum capacity. In  Fig. \ref{f8}, we can also see that the user rate $R_2$ increases monotonously with $\gamma_1$, but $R_1$ and $R_3$ decrease monotonously with $\gamma_1$. Similarly, the user rate $R_3$ increases monotonously with $\gamma_2$, but $R_1$ and $R_2$ decrease monotonously with $\gamma_2$. In Fig. \ref{f9}, the system capacity region is the polygonal consisted by the red lines, which is dominated by the red hexagonal face. The red points in Fig. \ref{f9} are the achievable points of the Iterative LMMSE detection. It shows that as we change the values of $\gamma_1$ and $\gamma_2$, the achievable points of the Iterative LMMSE detection can reach any point on the dominate hexagonal face. Therefore, for the three-user MIMO-NOMA systems the Iterative LMMSE detection can also achieve the whole system capacity region, i.e., the Iterative LMMSE detection is an optimal detection. In addition, we can see that the achievable rates of TDMA and OFDMA are strictly smaller than that of the Iterative LMMSE NOMA systems. It should be noted that the results in this paper can also apply to the case (like Fig. \ref{f8}) that the number of users is larger than the number of antennas, i.e., $N_u>N_r$.\vspace{-0.4cm}

\section{Conclusion}
We studied an Iterative LMMSE detector for MIMO-NOMA systems, which has a low-complexity as the distributed calculations replace the overall processing. The achievable rate region of the Iterative LMMSE detector has been analysed for the symmetric and asymmetric MIMO-NOMA systems. For the symmetric MIMO-NOMA systems, it is proved that the Iterative LMMSE detector is capacity achieving, and for the asymmetric MIMO-NOMA systems, we prove that the Iterative LMMSE detector is sum capacity achieving. In addition, it is showed that with the carefully designed Iterative LMMSE detector, all the maximal extreme points in the capacity region of asymmetric MIMO-NOMA systems are achievable, and the whole capacity regions of two-user and three-user asymmetric systems are also achievable.

\appendices
\section{An alternative Proof of (\ref{e42}) }
An alternative proof of (\ref{e42}) can be derived based on (\ref{e36}) at the estimator directly as follows.
\begin{eqnarray}\label{e43}
R_i &\mathop \leq \limits^{(a)}& \int \limits_{\phi_i( \mathbf{v}_{\bar{\mathbf{x}}} )=0} ^{\phi_i(\mathbf{v}_{\bar{\mathbf{x}}})=\infty}  v_{\hat{x}_i}(\mathbf{v}_{\bar{\mathbf{x}}}) d \phi_i(\mathbf{v}_{\bar{\mathbf{x}}})\nonumber\\
&\mathop \approx \limits^{(b)}& \int\limits_{\phi(v)=0}^{\phi(v)=\infty}  v_{\hat{x}}(v) d \phi(v)\nonumber\\
&\mathop  = \limits^{(c)}&\int\limits_{\phi(v)=0}^{\phi(v)=\phi(1)}  \left( 1+ \phi(v) \right)^{-1} d \phi(v) + \int\limits_{\phi(v)=\phi(1)}^{\phi(v)=\phi(0)}  v_{\hat{x}}(v) d \phi(v)  \nonumber\\
&=&- \log v_{\hat{x}}(v=1) + \int\limits_{v=1}^{v=0}  v_{\hat{x}}(v) d \left( v_{\hat{x}}(v)^{-1} - v^{-1}\right)  \nonumber\\
&=&\int\limits_{v=1}^{v=0}  { v_{\hat{x}}(v) d {v_{\hat{x}}(v)}^{-1}} - \int\limits_{v=1}^{v=0}  {v_{\hat{x}}(v)d{v^{-1} }} - \log v_{\hat{x}}(v=1)  \nonumber\\
&=& { \frac{1}{N_u}\log \det\left( \mathbf{I}_{N_r} + \frac{w^2}{\sigma _n^{ 2}}\mathbf{H}\mathbf{H}^H \right) } \nonumber\\
&=& R.
\end{eqnarray}
The inequality $(a)$ is based on the area property (\ref{e36}) and the equality holds if and only if there exists that code whose transfer function satisfies the matching condition. The approximation $(b)$ come from (\ref{e37}). Equation $(c)$ is based on the fact that the value region $[0,\phi(1)]$ of $\phi(v)$ is corresponding to a single value $v_i=1$, and $v_{\hat{x}}(v)=0$ if $\phi(v)>\phi(0)$. The following equations in (\ref{e43}) are similar with that of the (\ref{e42}).

\section{An alternative Proof of (\ref{e47}) }
The achievable rate of user $i$ can also be derived based on (\ref{e36}) at the estimator directly as follows.
\begin{eqnarray}\label{e48}
R_i&\mathop \leq \limits^{(a)} &\int\limits_{\phi_i(\mathbf{v}_{\bar{\mathbf{x}}})=0} ^{\phi_i(\mathbf{v}_{\bar{\mathbf{x}}})=\infty}  v_{\hat{x}_i}(\mathbf{v}_{\bar{\mathbf{x}}}) d \phi_i(\mathbf{v}_{\bar{\mathbf{x}}})\nonumber\\
&\mathop = \limits^{(b)}&\int\limits_{\phi_i(v_i=1)}^{\phi_i(v_i=0)}  {v_{\hat{x}_i}(v_i)d{\phi_i(v_i) }} + \int\limits_{0}^{\phi_i(v_i=1)} {(1+\phi_i(v_i))^{-1}d\phi_i(v_i)}  \nonumber\\
&\mathop = \limits^{(c)}&\int\limits_{v_i=1}^{v_i=0}  {v_{\hat{x}_i}(v_i)d\left( v_{\hat{x}_i}(v_i)^{-1}-v_i^{-1}\right)  } - \log v_{\hat{x}_i}({v}_{i}=1)  \nonumber\\
&\mathop = \limits^{(d)}& - \int\limits_{v_1=1}^{v_1=0} { \gamma_i^{-1} \left[\mathbf{V}_{\hat{\mathbf{x}}}(v_1)\right]_{i,i} dv_1^{-1}}
-\mathop {\lim }\limits_{v_1 \to 0} \; \log(\gamma_iv_1),
\end{eqnarray}

Similarly, the inequality $(a)$ is derived by the matching condition (\ref{em1})$\sim$(\ref{em3}) and the equality holds if and only if there exists that code whose transfer function satisfies the matching condition. Equations $(b)$ and $(c)$ are based on (\ref{evall})$\sim$(\ref{ev_1}).

\section{The Codes Existence of Lemma 1}
From the ``\emph{Property of SCM Codes}", we can see that there exist such $n$-layer SCM codes whose transfer function satisfies (i)$\sim$(iv) when $n$ is large enough. Therefore, it only needs to check the matched transfer function meets the conditions (i)$\sim$(iv) in order to show the existence of such code. It is easy to see that conditions (i) and (iv) are always satisfied by (\ref{em1}) and (\ref{em2}) respectively. From (\ref{ephi})$\sim$(\ref{em3}), we can see that $\psi_i(\rho_i)$ is continuous and differentiable in $[0,\infty)$ except at $\rho_i=\phi_i(0)$ and $\rho_i=\phi_i(1)$. Thus, Condition (iii) is satisfied. To show the monotonicity of the transfer function, we first rewritten (\ref{e38}) by the random matrix theorem as
\begin{eqnarray}\label{er2}
\phi_i(v_i)&=& { {{{\left[ {v_i - {v_i^2}\frac{{{w^2}}}{{\sigma _n^2}}\mathbf{h}_i^H{{\left( {{{\bf{I}}_{{N_r}}} + \frac{{{w^2}v_i}}{{\sigma _n^2}}{\bf{H}}{{\bf{H}}^H}} \right)}^{ - 1}}{\mathbf{h}_i}} \right]}^{ - 1}} - {v_i^{ - 1}}} } \nonumber \\
 &=&
  {\left[ {{{\left( {\frac{{{w^2}}}{{\sigma _n^2}}\mathbf{h}_i^H{{\left( {{v_i^{ - 1}}{{\bf{I}}_{{N_r}}} + \frac{{{w^2}}}{{\sigma _n^2}}{\bf{H}}{{\bf{H}}^H}} \right)}^{ - 1}}{\mathbf{h}_i}} \right)}^{ - 1}} - 1} \right]}^{ - 1} \nonumber \\
 &=&  {{1 \mathord{\left/
 {\vphantom {1 {\left( {f_i^{ - 1}(v_i) - 1} \right)}}} \right.
 \kern-\nulldelimiterspace} {\left( {f_i^{ - 1}(v_i) - 1} \right)}}},
\end{eqnarray}
where $f_i(v_i)={\frac{{{w^2}}}{{\sigma _n^2}}\mathbf{h}_i^H{{\left( {{v_i^{ - 1}}{{\bf{I}}_{{N_r}}} + \frac{{{w^2}}}{{\sigma _n^2}}{\bf{H}}{{\bf{H}}^H}} \right)}^{ - 1}}{\mathbf{h}_i}}$. It is easy to check that $f_i(v_i)$ is a decreasing function with respect to $v_i$, and $\phi_i(v_i)$ is thus a decreasing function of $v$. With the definition of $\psi(\rho)$ from (\ref{em1})$\sim$(\ref{em3}), we then see that $\psi_i(\rho_i)$ is a monotonically decreasing function in $[0,\infty)$. Therefore, the matched transfer function can be constructed by the SCM code, i.e., there exists such codes that satisfy the matching condition.

\section{The Existence of Infinite Integral (\ref{lemma1})}
With (\ref{lemma1}), we have
\begin{eqnarray}
{R_i} &=&  - \int\limits_{{v_1} = 1}^{{v_1} = 0} {\gamma _i^{ - 1}{{\left[ {{{\bf{V}}_{{\bf{\hat x}}}}({v_1})} \right]}_{i,i}}dv_1^{ - 1}}  - \mathop {\lim }\limits_{{v_1} \to 0} \;\log ({\gamma _i}{v_1})\nonumber\\
  &=&  - \int\limits_0^\infty  {{{\left[ {{{\left( {{\mathbf{A}_{\bm{\gamma}} } + s{{\bf{I}}_{{N_u}}}} \right)}^{ - 1}}} \right]}_{i,i}}ds}  - \mathop {\lim }\limits_{s \to \infty } \;\log ({\gamma _i}s^{-1}), \;\; s=v_1^{-1}, \mathbf{A}_{\bm{\gamma} } =\bm{\Lambda} _{\bm{\gamma} }^{1/2}\left( {\sigma _n^{ - 2}{{{\bf{H'}}}^H}{\bf{H'}} + {{\bf{I}}_{{N_u}}}} \right)\bm{\Lambda} _{\bm{\gamma} }^{1/2}
  \nonumber\\
  &=&  - \int\limits_0^\infty  {{\mathbf{u}_i}^H{{\left( {{\bm{\Lambda} _{{A_{\bm{\gamma} } }}} + s{{\bf{I}}_{{N_u}}}} \right)}^{ - 1}}{\mathbf{u}_i}ds}  - \mathop {\lim }\limits_{s \to \infty } \;\log ({\gamma _i}s^{-1}), \quad \mathbf{A}_{\bm{\gamma} } = \mathbf{U}^H\bm{\Lambda}_{\mathbf{A}_{\bm{\gamma} }} \mathbf{U} , \mathbf{u}_i\to i\mathrm{th}\;column\; of\; \mathbf{U} \nonumber\\
 &=&   - \int\limits_0^\infty  {\sum\limits_{j = 1}^{{N_u}} {{{\left\| {{u_{ij}}} \right\|}^2}} {{\left( {{\lambda _{{{\mathbf{A}_{\bm{\gamma}, }}}j}} + s} \right)}^{ - 1}}ds}  - \mathop {\lim }\limits_{s \to \infty } \;\log ({\gamma _i}{s^{ - 1}}),\quad {\lambda _{{{\mathbf{A}_{\bm{\gamma} }},}j}}\to i\mathrm{th}\; diagonal\; element \; of\;{\bm{\Lambda} _{{A_{\bm{\gamma} } }}} \nonumber\\
  &=&  \sum\limits_{j = 1}^{{N_u}} {{{\left\| {{u_{ij}}} \right\|}^2}} \log \left( {{\lambda _{{{\mathbf{A}_{\bm{\gamma} }},}j}}} \right) - \log ({\gamma _i})
\end{eqnarray}
Thus, we show the existence of the infinite integral (\ref{e47}) or (\ref{e48}), i.e., $R_i$ has a finite value.


\newpage
\begin{thebibliography}{h}
\bibitem{METIS}
METIS, ``Proposed solutions for new radio access," \emph{Mobile and wireless
communications enablers for the 2020 information society (METIS),}
Deliverable D.2.4, Feb. 2015.

\bibitem{5GWhitepaper}
 ``5G radio access: requirements, concepts and technologies," \emph{NTT DOCOMO, Inc., Tokyo, Japan, 5G Whitepaper}, Jul. 2014.

\bibitem{Kim2015}
B. Kim and W. Chung, ``Uplink NOMA with Multi-Antenna," \emph{in Proc. of
IEEE VTC 2015-Spring, Scotland, UK}, 2015.
\bibitem{Chen2015}
S. Chen, K. Peng and H. Jin, ``A suboptimal scheme for uplink NOMA in 5G systems," \emph{in Proc. of IEEE International Wireless Communications and Mobile Computing Conference (IWCMC)}, Aug. 2015.

\bibitem{Al-Imari2014}
M. Al-Imari, P. Xiao, M. A. Imran, and R. Tafazolli, ``Uplink non-orthogonal multiple access for 5g wireless networks," in \emph{Proc. of the 11th Int. Symp. on Wireless Commun. Systems (ISWCS), Barcelona, Spain}, Aug 2014, pp. 781-785.

 \bibitem{Saito2013}
Y. Saito, Y. Kishiyama, A. Benjebbour, T. Nakamura, A. Li, and K. Higuchi, ``Non-orthogonal multiple access (NOMA) for cellular future radio access," in \emph{Proc. IEEE Vehicular Technology Conference, Dresden, Germany}, Jun. 2013.
\bibitem{Ding2014}
Z. Ding, Z. Yang, P. Fan, and H. V. Poor, ``On the performance of non-orthogonal multiple access in 5G systems with randomly deployed users," \emph{IEEE Signal Process. Letters}, vol. 21, no. 12, pp. 1501-1505, Dec 2014.

\bibitem{Ding2015}
Z. Ding, M. Peng, and H. V. Poor, ``Cooperative non-orthogonal multiple
access in 5G systems," \emph{IEEE Commun. Lett.}, vol. 19, no. 8, pp. 1462-
1465, Aug. 2015.

\bibitem{Ding20161}
Z. Yang, Z. Ding, P. Fan, and G. K. Karagiannidis, ``On the Performance of Non-orthogonal Multiple Access Systems With Partial Channel Information," \emph{IEEE Trans. Commun.}, vol. 64, no. 2, pp. 654-
667, Feb. 2015.

\bibitem{Ding20162}
Z. Ding, F. Adachi, and H. V. Poor, ``The application of MIMO to
non-orthogonal multiple access," \emph{IEEE Trans. Wireless Commun.}, 2016,
submitted for publication.

\bibitem{Argas2013}
D. Argas, D. Gozalvez, D. Gomez-Barquero, and N. Cardona, ``MIMO for DVB-NGH, the next generation mobile TV broadcasting," \emph{IEEE Commun. Mag.}, vol. 51, no. 7, pp. 130-137, Jul. 2013.
\bibitem{Rusek2013}
F. Rusek, D. Persson, B. K. Lau, E. G. Larsson, T. L. Marzetta, O. Edfors, and F. Tufvesson, ``Scaling up MIMO: Opportunities and challenges with very large arrays," \emph{IEEE Signal Process. Mag.}, vol. 30, no. 1, pp. 40-60, Jan. 2013.
\bibitem{biglieri2007}
E. Biglieri, R. Calderbank, A. Constantinides, A. Goldsmith, A. Paulraj, and H. V. Poor, \emph{MIMO Wireless Communications}. Cambridge University Press, Cambridge, 2007.
\bibitem{Marzetta2010}
T. L. Marzetta, ``Noncooperative cellular wireless with unlimited numbers of base station antennas," \emph{IEEE Trans. Wireless Commun.}, vol. 9, no. 11, pp. 3590-3600, Nov. 2010.
%



\bibitem{Micciancio2001}
D. Micciancio, ``The hardness of the closest vector problem with
preprocessing," \emph{IEEE Transactions on Information Theory}, vol. 47, no. 3, pp. 1212-1215, Mar. 2001.
\bibitem{verdu1984_1}
S. Verd\'{u}, ``Optimum multi-user signal detection," Ph.D. dissertation, Department of Electrical and Computer Engineering, University of Illinois at Urbana-Champaign, Urbana, IL, Aug. 1984.
\bibitem{verdu1984_2}
S. Verd\'{u} and H. V. Poor, ``Backward, forward and backward-forward dynamic programming models under commutativity conditions," in P\emph{roc. the 23rd IEEE Conf. on Decision and Control (CDC'84)}, Las Vegas, NV, Dec. 1984, pp. 1081-1086.
\bibitem{verdu1986}
S. Verd\'{u}, ``Optimum sequence detection of asynchronous multipleaccess communications," in \emph{Abstr. IEEE International Symposium on Information Theory (ISIT'83)}, St. Jovite, Canada, Sep. 1983, p. 80.
\bibitem{verdu1987}
S. Verd\'{u} and H. V. Poor, ``Abstract dynamic programming models under commutativity conditions," \emph{SIAM Journal on Control and Optimization}, vol. 25, no. 4, pp. 990-1006, Jul. 1987.


\bibitem{tse2005}
Tse David and Pramod Viswanath, \emph{Fundamentals of wireless communication.} Cambridge university press, 2005.
\bibitem{Loeliger2004}
H. A. Loeliger, ``An introduction to factor graphs," \emph{IEEE Signal Processing Mag.}, pp. 28-41, Jan. 2004.
\bibitem{Loeliger2002}
H. A. Loeliger, ``Least squares and Kalman filtering on Forney graphs," \emph{in Codes, Graphs, and Systems, (festschrift in honor of David Forney on the occasion of his 60th birthday), R.E. Blahut and R. Koetter, eds., Kluwer}, 2002, pp. 113-135.
\bibitem{Loeliger2006}
H. A. Loeliger, J. Hu, S. Korl, Q. Guo and L. Ping, ``Gaussian message passing on linear models: an update," \emph{Int. Symp. on Turbo codes and Related Topics}, Apr. 2006.

\bibitem{Axelsson1994}
O. Axelsson, \emph{Iterative Solution Methods.} Cambridge, UK: Cambridge University Press, 1994.
\bibitem{Bertsekas1989}
D. P. Bertsekas and J. N. Tsitsiklis, \emph{Parallel and Distributed Calculation. Numerical Methods.} Prentice Hall, 1989.

\bibitem{Gao2014}
X. Gao, L. Dai, C. Yuen, and Y. Zhang, ``Low-Complexity MMSE Signal Detection Based on Richardson Method for Large-Scale MIMO Systems," \emph{in IEEE 80th Vehicular Technology Conference (VTC Fall)}, Sept. 2014, pp. 1-5.
\bibitem{Lei2015}
Lei Liu, Yuen Chau, Yong Liang Guan, Ying Li and Yuping Su, ``A Low-Complexity Gaussian Message Passing Iterative Detection for Massive MU-MIMO Systems," in \emph{Proc IEEE International Conference on Information, Communications and Signal Processing (ICICS)}, Singpore, Dec. 2015.
\bibitem{andrea2005}
A. Montanari, B. Prabhakar, and David Tse, ``Belief Propagation Based Multi-User Detection," \emph{Proceedings}, Vol. 43, 2005.
\bibitem{Roy2001}
P. Rusmevichientong and B. Van Roy, ``An analysis of belief propagation on the turbo decoding graph with Gaussian densities", \emph{IEEE Trans. Inform. Theory}, vol. 47, pp.745-765, 2001.

\bibitem{yoon2014}
S. Yoon and C. Chae, ``Low-Complexity MIMO Detection Based on Belief Propagation Over Pairwise Graphs," \emph{IEEE Trans. on TVT}, vol. 63, no. 5, pp. 2363-2377, 2014.


\bibitem{Cover2006}
T. M. Cover and J. A. Thomas, \emph{Elements of Information Theory-Second Edition}. New York: Wiley, 2006.
\bibitem{Gamal2012}
A. E. Gamal and Young-Han Kim. \emph{Network information theory}. Cambridge University Press, January 2012.
\bibitem{verdu1998}
S. Verd\'{u}, \emph{Multiuser Detection}. Cambridge, UK: Cambridge University Press, 1998.
\bibitem{Golden1999}
G. D. Golden, G. J. Foschini, R. A. Valenzuela, and P. W. Wolniansky, ``Detection algorithm and initial laboratory results using V-BLAST spacetime communication architecture," \emph{Electron. Lett.}, vol. 35, no. 1, pp. 14-16, January 1999.
\bibitem{Wang1999}
X. Wang and H. Poor, ``Iterative (turbo) soft interference cancellation and decoding for coded CDMA," \emph{IEEE Trans. Commun.}, vol. 47, no. 7, pp. 1046-1061, Jul. 1999.
\bibitem{Choi2000}
W. J. Choi, K. W. Cheong, and J. M. Cioffi, ``Iterative soft interference cancellation for multiple antenna systems," in \emph{Proc. IEEE Wireless Commun. Netw. Conf. (WCNC)}, 2000, vol. 1, pp. 304-309.
\bibitem{Studer2011}
C. Studer, S. Fateh, and D. Seethaler, ``ASIC implementation of soft-input soft-output MIMO detection using MMSE parallel interference cancellation," \emph{IEEE J. Solid-State Circuits}, vol. 46, no. 7, pp. 1754-1765, Jul. 2011.
\bibitem{Ping2003_1}
P. Li, L. Liu, K. Y. Wu, and W. K. Leung, ``Interleave-division multiple-access (IDMA) communications," in \emph{Proc. Int. Symp. Turbo Codes Related Topics, Brest, France, Sept. 2003}, pp. 173-180.
\bibitem{Ping2004}
P. Wang, J. Xiao, and P. Li, ``Comparison of Orthogonal and Non-Orthogonal Approaches to Future Wireless Cellular Systems," \emph{IEEE Vehicular Technology Magazine}, vol. 1, no. 3, pp. 4-11, Sept. 2006.
\bibitem{Ping2004_2}
L. Ping, L. Liu, K. Y.Wu, and W. K. Leung, ``Approaching the Capacity of Multiple Access Channels Using Interleaved Low-Rate Codes," \emph{IEEE Commun. letters}, vol. 8, no. 1, pp. 4-6, Jan. 2004.
\bibitem{Guo2008}
Q. Guo and L. Ping, ``LMMSE turbo equalization based on factor graphs," \emph{IEEE J. Sel. Areas Commun.}, vol. 26, no. 2, pp. 311-319, 2008.
\bibitem{wu2014}
S. Wu, L. Kuang, Z. Ni, J. Lu, D. DavidHuang, and Q. Guo, ``Low-Complexity Iterative Detection for Large-Scale Multiuser MIMO-OFDM Systems Using Approximate Message Passing," \emph{IEEE Selected Topics in Signal Processing}, vol. 8, no. 5, pp. 902-915, 2014.
\bibitem{Sanderovich2005}
A. Sanderovich, M. Peleg, and S. Shamai, ``LDPC coded MIMO multiple access with iterative joint decoding," \emph{IEEE Trans. Inf. Theory}, vol. 51, no. 4, pp. 1437-1450, Apr. 2005.
\bibitem{Caire2004}
G. Caire, R. Muller, and T. Tanaka, ``Iterative multiuser joint decoding: Optimal power allocation and low-complexity implementation," \emph{IEEE Trans. Inf. Theory}, vol. 50, no. 9, pp. 1950-1973, Sep. 2004.
\bibitem{Yuan2014}
X. Yuan, L. Ping, C. Xu and A. Kavcic, ``Achievable Rates of MIMO Systems With Linear Precoding and Iterative LMMSE Detection," \emph{IEEE Trans. Inf. Theory}, vol. 60, no.11, pp. 7073-7089, Oct. 2014.
\bibitem{Bhattad2007}
K. Bhattad and K. R. Narayanan, ``An MSE-based transfer chart for
analyzing iterative decoding schemes using a Gaussian approximation," \emph{IEEE Trans. Inf. Theory}, vol. 53, no. 1, pp. 22-38, Jan. 2007.
\bibitem{Guo2005}
D. Guo, S. Shamai, and S. Verd\'{u}, ``Mutual information and minimum
mean-square error in Gaussian channels," \emph{IEEE Trans. Inf. Theory}, vol. 51, no. 4, pp. 1261-1282, Apr. 2005.
\bibitem{Ashikhmin2004}
A. Ashikhmin, G. Kramer, and S. ten Brink, ``Extrinsic information
transfer functions: Model and erasure channel properties," \emph{IEEE Trans. Inf. Theory}, vol. 50, no. 11, pp. 2657-2673, Nov. 2004.
\bibitem{Brink2001}
S. ten Brink, ``Convergence behavior of iteratively decoded parallel concatenated codes," \emph{IEEE Trans. Commun.}, vol. 49, no. 10, pp. 1727-1737, Oct. 2001.
\bibitem{Andrews2007}
K. S. Andrews, D. Divsalar, S. Dolinar, J. Hamkins, C. R. Jones, F. Pollara, ``The Development of Turbo and LDPC Codes for Deep Space Applications," \emph{Proceedings of the IEEE}, Vol. 95, No. 11, Nov. 2007.

\bibitem{Richardson2001}
T. J. Richardson and R. L. Urbanke, ``The capacity of low-density paritycheck codes under message-passing decoding," \emph{IEEE Trans. Inf. Theory}, vol. 47, no. 2, pp. 599-618, Feb. 2001.
\bibitem{Han1979}
T. S. Han, ``The capacity region of general multiple-access channel with certain correlated sources," Inf. Control, vol. 40, no. 1, pp. 37-60, 1979.
\bibitem{Kay1993}
S. Kay, \emph{Fundamentals of Statistical Signal Processing: Estimation Theory.} Upper Saddle River, NJ, USA: Prentice-Hall, 1993.
\bibitem{Poor1997}
H. V. Poor and S. Verd\'{u}, ``Probability of error in MMSE multiuser detection," \emph{IEEE Trans. Inf. Theory}, vol. IT-43, no. 3, pp. 835-847, May 1997.
\bibitem{Guo2011}
D. Guo, Y. Wu, S. Shamai, and S. Verd\'{u}, ``Estimation in Gaussian noise: Properties of the minimum mean-square error," \emph{IEEE Trans. Inf. Theory}, vol. 57, no. 4, pp. 2371-2385, Apr. 2011.

\bibitem{Wachsmann1999}
U. Wachsmann, R. F. H. Fischer, and J. B. Huber, ``Multilevel codes: Theoretical concepts and practical design rules," \emph{IEEE Trans. Inf. Theory}, vol. 45, no. 5, pp. 1361-1391, Jul. 1999.
\bibitem{Gadkari1999}
S. Gadkari and K. Rose, ``Time-division versus superposition coded modulation schemes for unequal error protection," \emph{IEEE Trans. Commun.}, vol. 47, no. 3, pp. 370-379, Mar. 1999.
\end{thebibliography}
\end{document}